\documentclass[12pt,a4paper]{article}
\usepackage{amsmath,amssymb}
\usepackage[nosort]{cite}
\usepackage{graphicx}
\usepackage{pdflscape}
\def\mmu{m_{\mu}}
\def\iep{i\varepsilon}
\newcommand{\amu}[1]{a^{\text{HVP}(#1)}_{\mu}}
\newcommand{\APF}[1]{A_{0}^{(#1)}}
\newcommand{\KG}[2]{K_{#1}^{(#2)}}
\newcommand{\KGt}[2]{\tilde{K}_{#1}^{(#2)}}
\def\nf{n_{f}}
\def\ds{\displaystyle}
\allowdisplaybreaks

\begin{document}

\begin{center}

{\Large\bf
Window quantities for the hadronic vacuum \\
polarization contributions to the muon anomalous \\
magnetic moment in spacelike and timelike domains

}

\vskip7.5mm

{\large A.V.~Nesterenko}

\vskip5mm

{\small\it Bogoliubov Laboratory of Theoretical Physics,
Joint Institute for Nuclear Research,\\
Dubna, 141980, Russian Federation}

\end{center}

\vskip2.5mm

\noindent
\centerline{\bf Abstract}

\vskip2.5mm

\centerline{\parbox[t]{150mm}{%
The relations between the window quantities for the hadronic vacuum
polarization contributions to the muon anomalous magnetic
moment~$a^{\text{HVP}}_{\mu}$ in spacelike and timelike domains are studied.
Two types of window functions (abrupt and smooth) as~well~as two kinds of
kinematic intervals (symmetric and asymmetric with respect to the
spacelike/timelike flip) are addressed. It~is shown that the window
quantities for~$a^{\text{HVP}}_{\mu}$ represented in terms of the hadronic
vacuum polarization function~$\bar\Pi(Q^2)$, the Adler function~$D(Q^2)$, and
the $R$--ratio of electron--positron annihilation into hadrons are mutually
equivalent only if the additional contributions due to the window edge
effects are properly taken into account and the explicit expressions for such
contributions are derived. The~obtained results enable one to
evaluate~$a^{\text{HVP}}_{\mu}$ by~making simultaneous use of the inputs for
functions~$\bar\Pi(Q^2)$, $D(Q^2)$, and~$R(s)$ at~various energies and an
example of such hybrid assessment is provided. The obtained results also
enable one to accurately compare the window quantities
for~$a^{\text{HVP}}_{\mu}$ based, e.g., on MUonE or lattice data with the
ones based on $R$--ratio data, even~if the window function covers different
kinematic ranges in spacelike and timelike domains.
\\[2.5mm]
}}

\vskip12mm

\section{Introduction}
\label{Sect:Intro}

The muon anomalous magnetic moment $a_{\mu} = (g_{\mu}-2)/2$ constitutes
a~most engaging issue of elementary particle physics, which encompasses all
the Standard Model interactions. The~measurements of~$a_{\mu}$ performed
within Brookhaven E821~\cite{Muong-2:2006rrc} and Fermilab
E989~\cite{Muong-2:2021ojo, Muong-2:2023cdq, Muong-2:2025xyk} experiments
have achieved an unprecedented accuracy, while a~number of planned projects,
such as MUonE at~CERN~\cite{CarloniCalame:2015obs, MUonE:2016hru,
MUonE:2019qlm, Banerjee:2020tdt}, E34~at~J--PARC~\cite{Abe:2019thb}, and
muEDM~at~PSI~\cite{Adelmann:2021udj, Renga:2024ypc}, are highly anticipated.
The~theoretical assessments of~$a_{\mu}$ (see a thorough
review~\cite{Aoyama:2020ynm} and its recent update~\cite{Aliberti:2025beg},
which is primarily based on Refs.~\cite{RBC:2018dos, Giusti:2019xct,
Borsanyi:2020mff, Lehner:2020crt, Wang:2022lkq, Aubin:2022hgm, Ce:2022kxy,
ExtendedTwistedMass:2022jpw, RBC:2023pvn, Kuberski:2024bcj,
Boccaletti:2024guq, Spiegel:2024dec, RBC:2024fic, Djukanovic:2024cmq,
ExtendedTwistedMass:2024nyi, MILC:2024ryz, FermilabLatticeHPQCD:2024ppc,
Keshavarzi:2019abf, DiLuzio:2024sps, Kurz:2014wya, Colangelo:2015ama,
Masjuan:2017tvw, Colangelo:2017fiz, Hoferichter:2018kwz, Eichmann:2019tjk,
Bijnens:2019ghy, Leutgeb:2019gbz, Cappiello:2019hwh, Masjuan:2020jsf,
Bijnens:2020xnl, Bijnens:2021jqo, Danilkin:2021icn, Stamen:2022uqh,
Leutgeb:2022lqw, Hoferichter:2023tgp, Hoferichter:2024fsj, Estrada:2024cfy,
Ludtke:2024ase, Deineka:2024mzt, Eichmann:2024glq, Bijnens:2024jgh,
Hoferichter:2024bae, Holz:2024diw, Cappiello:2025fyf, Colangelo:2014qya,
Blum:2019ugy, Chao:2021tvp, Chao:2022xzg, Blum:2023vlm, Fodor:2024jyn,
Aoyama:2012wk, Volkov:2019phy, Volkov:2024yzc, Aoyama:2024aly,
Parker:2018vye, Morel:2020dww, Fan:2022eto, Czarnecki:2002nt,
Gnendiger:2013pva, Hoferichter:2025yih}) have also attained an~impressive
precision, whereas a long--standing discrepancy of a few standard deviations
between the experimental measurements and theoretical data--driven
evaluations of~$a_{\mu}$ may eventually become an evidence for the existence
of a new fundamental physics beyond the Standard Model.

The uncertainty of theoretical estimation of~$a_{\mu}$ is largely dominated
by the hadronic contribution, which entails a tangled dynamics of coloured
fields at low energies, that remains beyond the applicability range of the
QCD~perturbative approach. Basically, the evaluation of the hadronic vacuum
polarization contributions to the muon anomalous magnetic
moment~$a^{\text{HVP}}_{\mu}$ can equivalently be performed within two
distinct methods. Specifically, the first (``spacelike'') method
represents~$a^{\text{HVP}}_{\mu}$ in terms of either the hadronic vacuum
polarization function~$\bar\Pi(Q^2)$ or the related Adler function~$D(Q^2)$,
whereas the second (``timelike'') method expresses~$a^{\text{HVP}}_{\mu}$ in
terms of the $R$--ratio of electron--positron annihilation into hadrons.
While none of the theoretical and experimental inputs for the foregoing
functions covers the entire kinematic interval required for the calculation
of~$a^{\text{HVP}}_{\mu}$, one commonly resorts~to the data--driven method,
that complements, e.g., low--energy data on the $R$--ratio with its
high--energy perturbative expression.

At the same time, an~approach, that combines various available inputs
from~different ranges (such~as lattice hadronic vacuum polarization function
at intermediate energies and $R$--ratio at low and high energies), was
suggested in Refs.~\cite{Bernecker:2011gh, Lehner:2017kuc, RBC:2018dos},
which prompted widespread use of the window quantities in the studies of the
muon anomalous magnetic moment. Specifically, since the window quantities
for~$a^{\text{HVP}}_{\mu}$ involve only a part of the kinematic range
(the~corresponding window function can be abrupt or smooth, which provides
additional control on the taken interval's endpoints), it~becomes possible to
evaluate them resorting to a single input, for~example, MUonE
measurements~\cite{CarloniCalame:2015obs, MUonE:2016hru, MUonE:2019qlm,
Banerjee:2020tdt}, lattice simulations (see~reviews~\cite{Meyer:2018til,
Gerardin:2020gpp, Wittig:2023pcl} as well as recent
results~\cite{Borsanyi:2020mff, Boccaletti:2024guq, Ce:2022kxy,
Kuberski:2024bcj, Djukanovic:2024cmq, RBC:2018dos, RBC:2023pvn, RBC:2024fic,
MILC:2024ryz, FermilabLatticeHPQCD:2024ppc}), or $R$--ratio measurements.
Additionally, the lattice~QCD has specific challenges in different regions
(such as discretization errors at short distances and finite lattice volume
effects at long distances), that can be handled by optimizing simulations
separately in those regions, see also Sect.~3.4 of
review~\cite{Aliberti:2025beg} and references therein for the details on this
matter.

The primary objective of this paper is to obtain the explicit relations
between the window quantities for the hadronic vacuum polarization
contributions to the muon anomalous magnetic moment in spacelike and timelike
domains for various types of window functions and covered energy ranges.

The layout of the paper is as follows. In~Sect.~\ref{Sect:Results} the
explicit expressions for the additional contributions to the window
quantities for~$a^{\text{HVP}}_{\mu}$ expressed in terms of the Adler
function and \mbox{$R$--ratio}, which appear due to the window edge effects,
are derived for several classes of window functions and kinematic intervals
and the obtained results are exemplified within dispersively improved
perturbation theory~(DPT)~\cite{Nesterenko:2013vja, Nesterenko:2014txa,
Nesterenko:2016pmx}. Section~\ref{Sect:Dsc} contains a~discussion, provides
an example of hybrid evaluation of~$a^{\text{HVP}}_{\mu}$, and reports
an~updated DPT~value of the muon anomalous magnetic moment.
Section~\ref{Sect:Concl} summarizes the basic results.

\section{Results}
\label{Sect:Results}

This study generally continues those of Refs.~\cite{Nesterenko:2021byp,
Nesterenko:2023bdc}. All the definitions and notations employed hereinafter
can therefore be found in Sects.~2 of Refs.~\cite{Nesterenko:2021byp,
Nesterenko:2023bdc}.

\bigskip

The window quantities for the hadronic vacuum polarization contributions to
the muon anomalous magnetic moment~$a^{\text{HVP}}_{\mu,W}$ can be represented
in the \mbox{$\ell$--th}~order in the electromagnetic coupling as
\begin{subequations}
\label{AmuW}
\begin{align}
\label{AmuWP}
a^{\text{HVP}(\ell)}_{\mu,\Pi,W_{n}} & = \APF{\ell}\!\!\int\limits_{0}^{\infty}\!
\bar\Pi(Q^2) \KG{\Pi}{\ell}(Q^2) W_{n}(-Q^2)\frac{d Q^2}{4\mmu^2} =
\APF{\ell}\!\!\int\limits_{0}^{\infty}\!
\bar\Pi(4\zeta\mmu^2) \KGt{\Pi}{\ell}(\zeta) W_{n}(-4\zeta\mmu^2) d \zeta,
\\[1.25mm]
\label{AmuWD}
a^{\text{HVP}(\ell)}_{\mu,D,W_{n}} & = \APF{\ell}\!\!\int\limits_{0}^{\infty} \!
D(Q^2) \KG{D}{\ell}(Q^2) W_{n}(-Q^2)\frac{d Q^2}{4\mmu^2} =
\APF{\ell}\!\!\int\limits_{0}^{\infty} \!
D(4\zeta\mmu^2) \KGt{D}{\ell}(\zeta) W_{n}(-4\zeta\mmu^2)d \zeta, \\[1.25mm]
\label{AmuWR}
a^{\text{HVP}(\ell)}_{\mu,R,W_{n}} & = \APF{\ell}\!\!\int\limits_{s_{0}}^{\infty} \!
R(s) \KG{R}{\ell}(s) W_{n}(s)\frac{d s}{4\mmu^2} =
\APF{\ell}\!\!\int\limits_{\chi}^{\infty} \!
R(4\eta\mmu^2) \KGt{R}{\ell}(\eta) W_{n}(4\eta\mmu^2) d \eta.
\end{align}
\end{subequations}
In these equations~$\APF{\ell}$ is a constant prefactor, $\bar\Pi(Q^2) =
-\Pi(-Q^2)$ denotes the subtracted at zero hadronic vacuum polarization
function (see~Eq.~(1) of Ref.~\cite{Nesterenko:2023bdc}) with $\Pi(0)=0$
being assumed, \mbox{$Q^2 = -q^2 \ge 0$} and \mbox{$s = q^2 \ge 0$} stand,
respectively, for the spacelike and timelike kinematic variables,
\mbox{$\zeta = Q^2/(4\mmu^2)$} and~$\eta = s/(4\mmu^2)$ are the dimensionless
kinematic variables, \mbox{$\chi=s_{0}/(4\mmu^2)$}, $\KG{\Pi}{\ell}(Q^2)$,
$\KG{D}{\ell}(Q^2)$, and~$\KG{R}{\ell}(s)$ denote the corresponding kernel
functions, and~$W_{n}(q^2)$ stands for a~window function, which modulates the
integrands~(\ref{AmuW}). The~additional subscripts ``$\Pi$'', ``$D$'',
and~``$R$'' on the left--hand sides of Eqs.~(\ref{AmuW}) indicate that the
evaluated window quantity for~$a^{\text{HVP}}_{\mu}$ is represented in terms
of the hadronic vacuum polarization function~$\bar\Pi(Q^2)$, the Adler
function~$D(Q^2)$, and the $R$--ratio of electron--positron annihilation into
hadrons, respectively.

To obtain the relations between the window quantities
for~$a^{\text{HVP}}_{\mu}$ in spacelike and timelike domains it is convenient
to make use of the function
\begin{equation}
\label{FWDef}
F_{W_{n}}(q^2) =
-\frac{1}{4\mmu^2} \Pi(q^2) K_{R}(q^2) W_{n}(q^2) =
\frac{1}{4\mmu^2} \bar\Pi(-q^2) K_{R}(q^2) W_{n}(q^2).
\end{equation}
As~argued in, e.g., Ref.~\cite{Feynman:1973xc}, the function~$\Pi(q^2)$ has
the only cut (which will be shown by the red line in the contour plots
hereinafter) along the positive semiaxis of real~$q^2$ starting at the
hadronic production threshold~$q^2 \ge s_{0}$. In~turn, as~argued in, e.g.,
Ref.~\cite{Barbieri:1974nc}, the kernel function $K_{R}(q^2)$ possesses the
only cut (which will be shown by the orange line in the contour plots
hereinbelow) along the negative semiaxis of real~$q^2$ starting at the
origin~\mbox{$q^2 \le 0$}.

\subsection{Constant window function}
\label{Sect:WF1}

\begin{figure}[t]
\centerline{%
\includegraphics[height=70mm,clip]{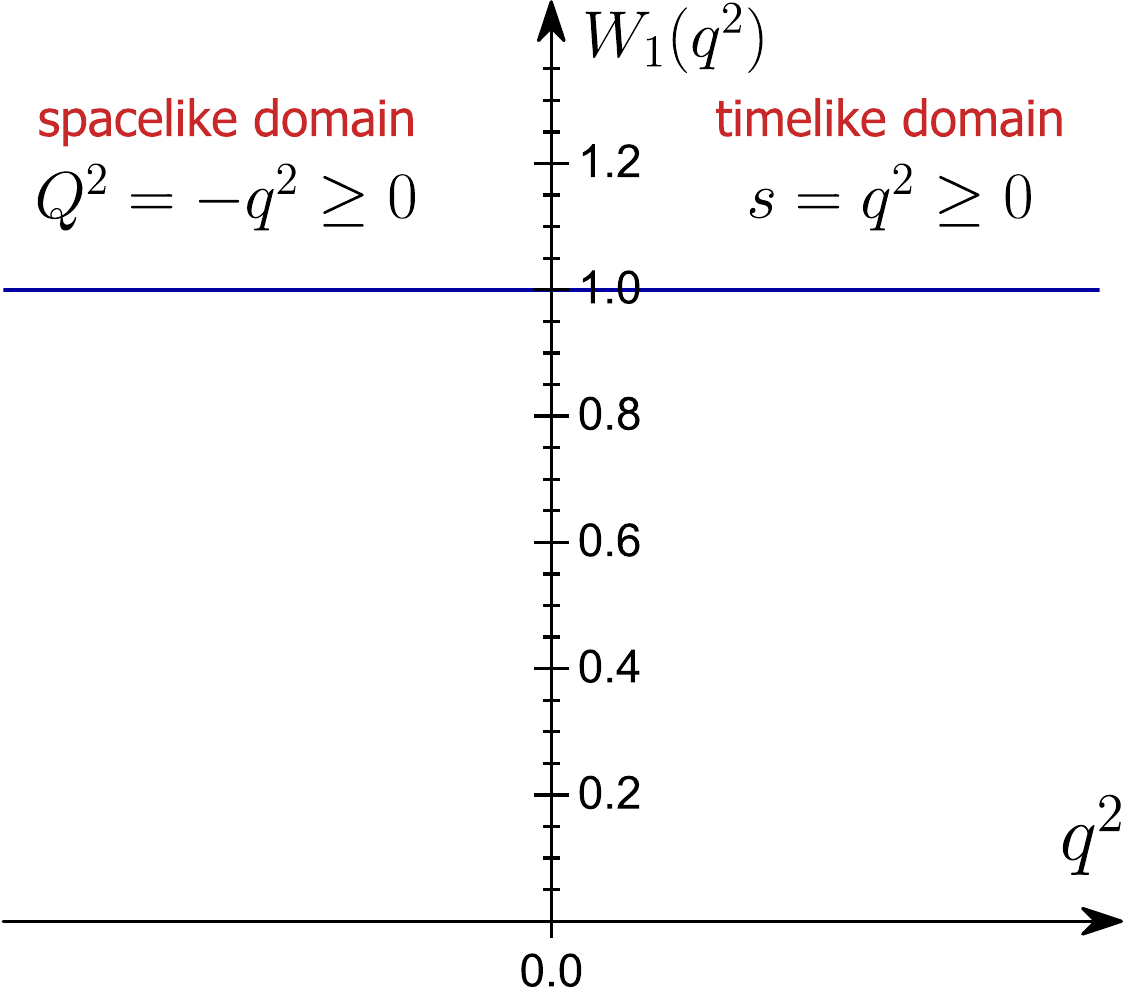}%
\hspace{10mm}%
\includegraphics[height=70mm,clip]{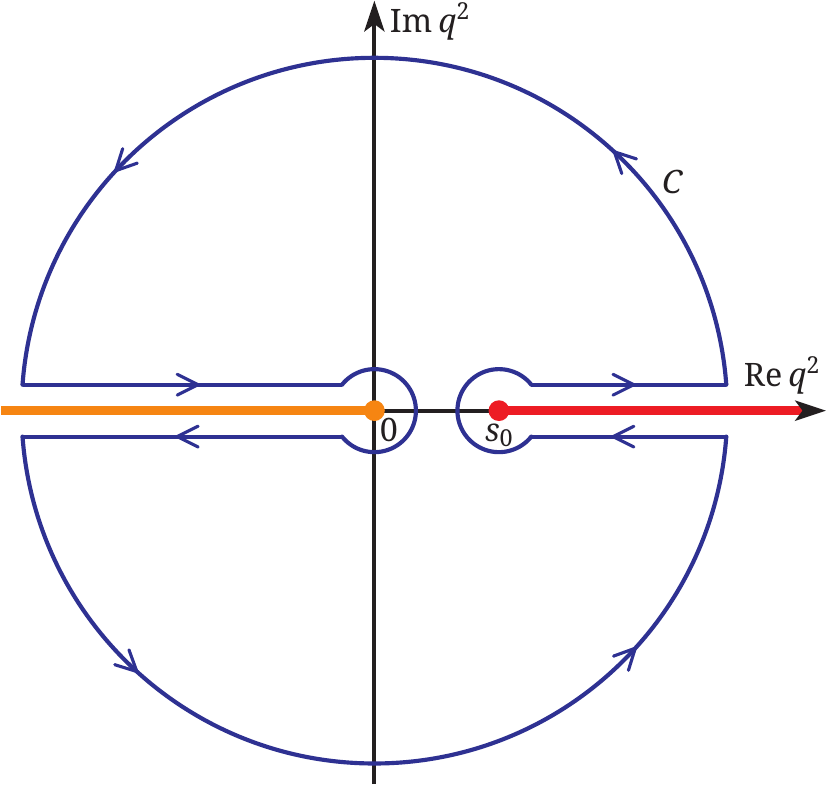}}
\caption{Left--hand plot: the constant window
function~$W_{1}(q^2)$~(\ref{DefW1}). Right--hand plot: the closed integration
contour~$C$ in the complex~$q^2$--plane in Eq.~(\ref{W1IntC1}).}
\label{Plot:WF1}
\end{figure}

To set up the stage for other cases, it is worthwhile to address the constant
window function (see left--hand plot of Fig.~\ref{Plot:WF1})
\begin{equation}
\label{DefW1}
W_{1}(q^2) = 1,
\end{equation}
that factually corresponds to the absence of the last multiplier on the
right--hand side of Eq.~(\ref{FWDef}), thereby making the study identical to
that of Sect.~3.1 of Ref.~\cite{Nesterenko:2021byp}. Specifically, the
integral of the product of the functions~$\bar\Pi(-q^2)$ and~$K_{R}(q^2)$
along the closed contour~$C$ displayed in the right--hand plot of
Fig.~\ref{Plot:WF1} vanish, namely
\begin{equation}
\label{W1IntC1}
\oint_{C}\! \bar\Pi(-q^2) K_{R}(q^2) \frac{d q^{2}}{4\mmu^2} = 0,
\end{equation}
that can be represented as
\begin{align}
\label{W1IntC2}
& \int\limits_{\infty-i\varepsilon}^{s_{0}-i\varepsilon}
\bar\Pi(-q^2) K_{R}(q^2) \frac{d q^2}{4\mmu^2}
+ \int\limits_{s_{0}+i\varepsilon}^{\infty+i\varepsilon}
\bar\Pi(-q^2) K_{R}(q^2) \frac{d q^2}{4\mmu^2}
+ \nonumber \\[1.5mm] &
+ \int\limits_{-\infty+i\varepsilon}^{i\varepsilon}
\bar\Pi(-q^2) K_{R}(q^2) \frac{d q^2}{4\mmu^2}
+ \int\limits_{-i\varepsilon}^{-\infty-i\varepsilon}
\bar\Pi(-q^2) K_{R}(q^2) \frac{d q^2}{4\mmu^2} = 0.
\end{align}
Upon rearrangement of terms, change of the integration variables, and overall
multiplication by a constant prefactor~$A_{0}$, Eq.~(\ref{W1IntC2}) takes the
form
\begin{align}
\label{W1IntC3}
& a^{\text{HVP}}_{\mu,\Pi} = a^{\text{HVP}}_{\mu,\Pi,W_{1}} =
A_{0}\!\int\limits_{0}^{\infty}\! \bar\Pi(Q^2) K_{\Pi}(Q^2) \frac{d Q^2}{4\mmu^2} =
\nonumber \\[1.5mm] = \, &
a^{\text{HVP}}_{\mu,R} = a^{\text{HVP}}_{\mu,R,W_{1}} =
A_{0}\!\int\limits_{s_{0}}^{\infty}\! R(s) K_{R}(s) \frac{d s}{4\mmu^2},
\end{align}
with the relations
\begin{equation}
\label{RDefP}
R(s) = \frac{1}{2 \pi i} \lim_{\varepsilon \to 0_{+}}
\Bigl[\Pi(s + \iep) - \Pi(s - \iep)\Bigr]^{\!}
\end{equation}
and
\begin{equation}
\label{KRelPR}
K_{\Pi}(Q^2) = \frac{1}{2 \pi i} \lim_{\varepsilon \to 0_{+}}
\Bigl[ K_{R}(-Q^2+i\varepsilon) - K_{R}(-Q^2-i\varepsilon)\Bigr],
\qquad
Q^2 \ge 0
\end{equation}
being employed (see, respectively, Eq.~(3) of Ref.~\cite{Nesterenko:2023bdc}
and Eq.~(21) of Ref.~\cite{Nesterenko:2021byp}). It~is~worthwhile to note
here that Eq.~(\ref{KRelPR}) has also been independently derived
in~Ref.~\cite{Balzani:2021del} in a way different from that of
Ref.~\cite{Nesterenko:2021byp}.

In other words, for the constant window function~(\ref{DefW1}) the
integration of~$F_{W_{1}}(q^2)$~(\ref{FWDef}) along the edges of its left cut
yields the first line of Eq.~(\ref{W1IntC3}), the integration
of~$F_{W_{1}}(q^2)$ along the edges of its right cut produces the second line
of Eq.~(\ref{W1IntC3}), whereas the integrals of~$F_{W_{1}}(q^2)$ along the
circles of infinitely large and infinitely small radii vanish, see
right--hand plot of~Fig.~\ref{Plot:WF1}.

In~turn, to obtain the relation between the window quantities
for~$a^{\text{HVP}}_{\mu}$ expressed in terms of functions~$\bar\Pi(Q^2)$
and~$D(Q^2)$, it is convenient to integrate Eq.~(\ref{AmuWD}) by parts,
specifically
\begin{equation}
\label{KRelPDaux1}
A_{0}\!\int\limits_{0}^{\infty}\! D(Q^2) K_{D}(Q^2) \frac{d Q^2}{4\mmu^2} =
A_{0}\, \bar\Pi(Q^2) K_{D}(Q^2) \frac{Q^2}{4\mmu^2}\, \Biggr|_{0}^{\infty}
+ A_{0}\!\int\limits_{0}^{\infty}\! \bar\Pi(Q^2) K_{\Pi}(Q^2) \frac{d Q^2}{4\mmu^2}.
\end{equation}
This equation employs the definition of the Adler
function~\cite{Adler:1974gd}
\begin{equation}
\label{GDR_DP}
D(Q^2) = -\,\frac{d\, \Pi(-Q^2)}{d \ln Q^2},
\end{equation}
and the relation between the kernel functions obtained in
Ref.~\cite{Nesterenko:2021byp}
\begin{equation}
\label{KRelPD}
K_{\Pi}(Q^2) = - \biggl[K_{D}(Q^2)
+ \frac{d\,K_{D}(Q^2)}{d\,\ln Q^2}\biggr]\!,
\qquad
Q^2 \ge 0.
\end{equation}
As~discussed in Ref.~\cite{Nesterenko:2021byp}, the first term on the
right--hand side of Eq.~(\ref{KRelPDaux1}) vanishes, hence
\begin{align}
\label{W1IntC4}
& a^{\text{HVP}}_{\mu,\Pi} = a^{\text{HVP}}_{\mu,\Pi,W_{1}} =
A_{0}\!\int\limits_{0}^{\infty}\! \bar\Pi(Q^2) K_{\Pi}(Q^2) \frac{d Q^2}{4\mmu^2} =
\nonumber \\[1.5mm] = \, &
a^{\text{HVP}}_{\mu,D} = a^{\text{HVP}}_{\mu,D,W_{1}} =
A_{0}\!\int\limits_{0}^{\infty}\! D(Q^2) K_{D}(Q^2) \frac{d Q^2}{4\mmu^2}.
\end{align}

Thus, Eqs.~(\ref{W1IntC3}) and~(\ref{W1IntC4}) imply that in the case of the
constant window function~(\ref{DefW1}) all three representations~(\ref{AmuW})
are equivalent to each other and yield the same result
for~$a^{\text{HVP}}_{\mu}$, namely
\begin{equation}
\label{AmuW1}
a^{\text{HVP}}_{\mu} =
a^{\text{HVP}}_{\mu,\Pi} =
a^{\text{HVP}}_{\mu,D} =
a^{\text{HVP}}_{\mu,R}.
\end{equation}
At the same time, it is necessary to emphasize here that Eq.~(\ref{AmuW1}) is
valid only for the continuous functions~$\bar\Pi(Q^2)$, $D(Q^2)$, and~$R(s)$,
which satisfy the complete set of dispersion relations given in, e.g.,
Chap.~1 of Ref.~\cite{Nesterenko:2016pmx} (see also Sects.~2.1 of
Refs.~\cite{Nesterenko:2021byp, Nesterenko:2023bdc}). As~for the case of the
piecewise continuous functions~$\bar\Pi(Q^2)$, $D(Q^2)$, and~$R(s)$ (that
takes place, e.g., in~the presence of the quark flavour thresholds), the
mutual equivalence of the representations of~$a^{\text{HVP}}_{\mu}$ in terms
of those three functions~(\ref{AmuW1}) is preserved only if the additional
contributions to~$a^{\text{HVP}}_{\mu,D}$ and~$a^{\text{HVP}}_{\mu,R}$ are
taken into account, see Ref.~\cite{Nesterenko:2023bdc} for the details.
Specifically, in~Sect.~3.1 of Ref.~\cite{Nesterenko:2023bdc} the explicit
expressions for such contributions are derived, while Sect.~3.2 thoroughly
illustrates this matter with the example of the dispersively improved
perturbation theory~(DPT)~\cite{Nesterenko:2013vja, Nesterenko:2014txa,
Nesterenko:2016pmx}, which will also be employed hereinafter to exemplify the
obtained results.

\subsection{Abrupt symmetric window function}
\label{Sect:WF2}

\begin{figure}[t]
\centerline{%
\includegraphics[height=70mm,clip]{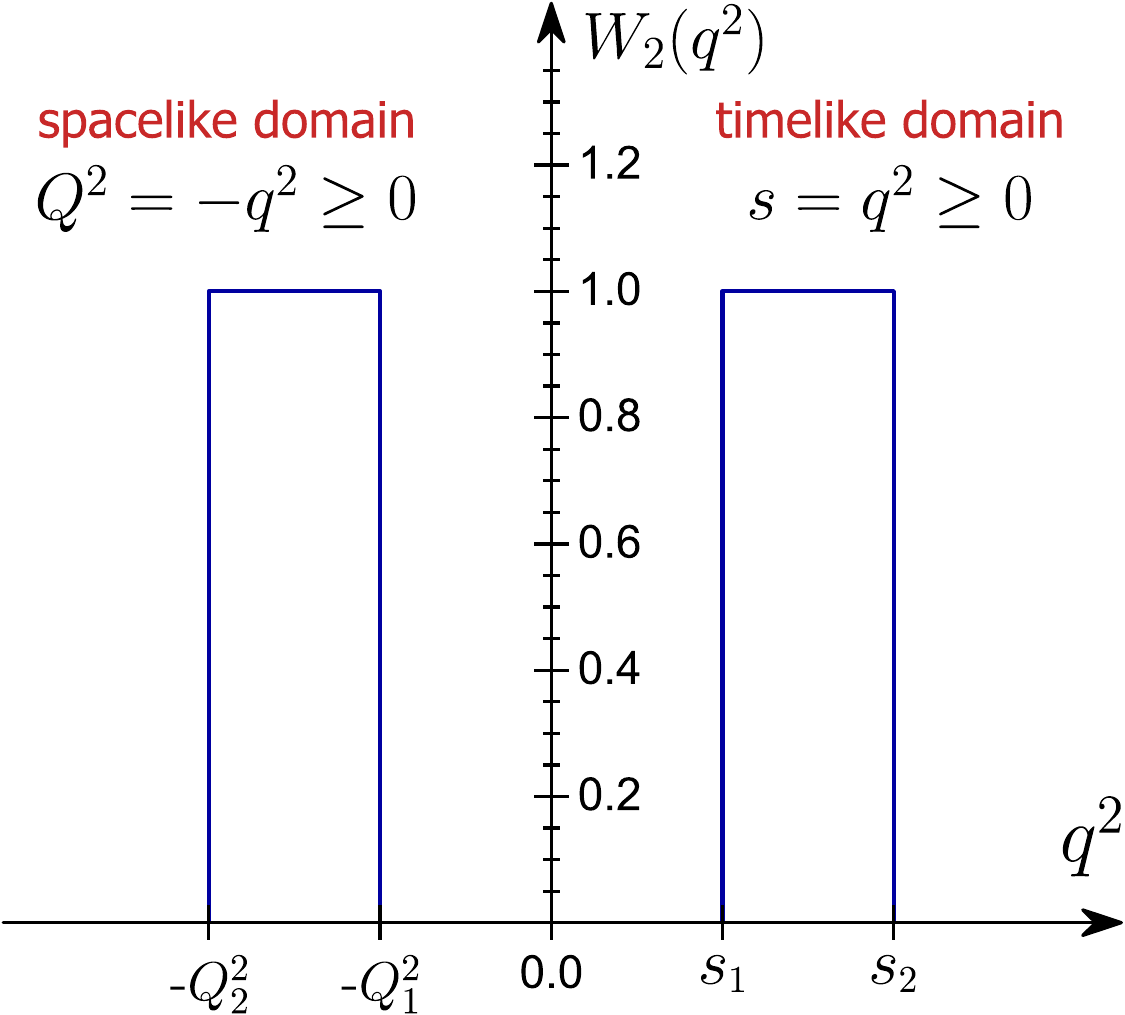}%
\hspace{10mm}%
\includegraphics[height=70mm,clip]{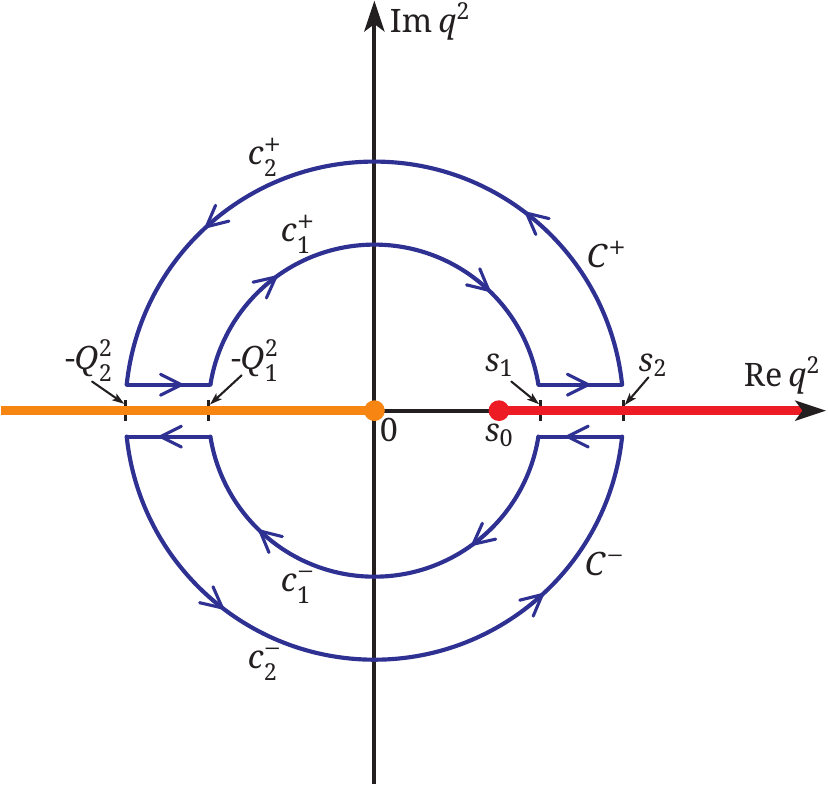}}
\caption{Left--hand plot: the abrupt symmetric window
function~$W_{2}(q^2)$~(\ref{DefW2}). Right--hand plot: the closed integration
contours in the complex~$q^2$--plane in Eq.~(\ref{IntC1C2}).}
\label{Plot:WF2}
\end{figure}

Let us proceed now to the abrupt window function (see left--hand plot of
Fig.~\ref{Plot:WF2}):
\begin{align}
\label{DefW2}
W_{2}(q^{2}) & =
\theta(q^{2}+Q_{2}^{2}) - \theta(q^{2}+Q_{1}^{2}) +
\theta(q^{2}-s_{1}) - \theta(q^{2}-s_{2}),
\nonumber \\[2mm] & \quad\;
\left|-Q_{1}^{2}\right| = s_{1},
\qquad
\left|-Q_{2}^{2}\right| = s_{2},
\end{align}
with $\theta(x)$ being the dimensionless piecewise Heaviside unit step
function
\begin{equation}
\label{DefUSF}
\theta(x) = \left\{
\begin{aligned}
1, \quad\; & x \ge 0, \\[0mm]
0, \quad\; & x < 0.   \\
\end{aligned}
\right.
\end{equation}
It is assumed hereinbelow that the variables~$Q_{i}$ and~$s_{i}$ ($i=1,2$)
take positive values and \mbox{$Q_{1}^{2} < Q_{2}^{2}$}, \mbox{$s_{1} <
s_{2}$}. The~window function $W_{2}(q^{2})$~(\ref{DefW2}) covers the coequal
kinematic intervals in spacelike ($Q_{1}^{2} \le Q^{2} \le Q_{2}^{2}$) and
timelike ($s_{1} \le s \le s_{2}$) domains and is therefore symmetric with
respect to the spacelike/timelike flip~$q^2 \leftrightarrow -q^2$.

Factually, the effect of the abrupt window
function~$W_{2}(q^{2})$~(\ref{DefW2}) on Eq.~(\ref{AmuW}) is the alternation
of the integration limits. Keeping this in mind, one can safely
omit~$W_{2}(q^{2})$ from Eq.~(\ref{FWDef}), that makes the study identical to
that of Sect.~3.1 of~Ref.~\cite{Nesterenko:2023bdc}. Specifically, the
integrals of the function~$F_{W_{2}}(q^2)$~(\ref{FWDef}) along each of the
closed contours~$C^{+}$ and~$C^{-}$ shown in the right--hand plot of
Fig.~\ref{Plot:WF2} vanish and, therefore, so does their sum, namely
\begin{equation}
\label{IntC1C2}
\oint_{C^{+}}\! F_{W_{2}}(q^2) d q^{2} +
\oint_{C^{-}}\! F_{W_{2}}(q^2) d q^{2} = 0.
\end{equation}
The rearrangement of terms, change of the integration variables, and overall
multiplication by a constant prefactor~$A_{0}$ cast this equation~to
\begin{equation}
\label{RelPRint1}
A_{0}\!\int\limits_{Q_{1}^{2}}^{Q_{2}^{2}} \!
\bar\Pi(Q^2) K_{\Pi}(Q^2) \frac{dQ^2}{4\mmu^2} =
A_{0}\!\int\limits_{s_{1}}^{s_{2}} \!
R(s) K_{R}(s) \frac{ds}{4\mmu^2}
- A_{0}\frac{\Delta F_{2}}{2\pi i},
\end{equation}
\begin{equation}
\label{DltFW2}
\Delta F_{2} =
\int_{c_{1}^{+}}\! F_{W_{2}}(q^2) d q^{2} +
\int_{c_{1}^{-}}\! F_{W_{2}}(q^2) d q^{2} +
\int_{c_{2}^{+}}\! F_{W_{2}}(q^2) d q^{2} +
\int_{c_{2}^{-}}\! F_{W_{2}}(q^2) d q^{2},
\end{equation}
with the relations~(\ref{RDefP}) and~(\ref{KRelPR}) being used. Then, it is
convenient to perform the integration in~Eq.~(\ref{DltFW2}) in the polar
coordinates, that ultimately yields the relation between the
spacelike~(\ref{AmuWP}) and timelike~(\ref{AmuWR}) window quantities (see
Eq.~(36) of Ref.~\cite{Nesterenko:2023bdc})
\begin{equation}
\label{RelPRint2}
a^{\text{HVP}}_{\mu,\Pi,W_{2}} =
a^{\text{HVP}}_{\mu,R,W_{2}} +
\Delta a^{\text{HVP}}_{\mu,R,W_{2}},
\end{equation}
where
\begin{equation}
\label{DltAmuRW2}
\Delta a^{\text{HVP}}_{\mu,R,W_{2}} =
- A_{0}\Bigl[ T(s_{2}) - T(s_{1}) \Bigr],
\end{equation}
\begin{equation}
\label{THdef}
T(q^{2}) = \frac{1}{2\pi}\!
\left[ \int\limits_{\varepsilon}^{\pi-\varepsilon}\!
H(q^{2},\varphi) d\varphi
+
\int\limits_{\pi+\varepsilon}^{2\pi-\varepsilon}\!
H(q^{2},\varphi) d\varphi \right]\!\!,
\quad
H(q^{2},\varphi) =
\bar\Pi(-q^{2}e^{i\varphi})
G_{R}\!\left(\frac{q^2}{4\mmu^2}e^{i\varphi}\!\right)\!,
\end{equation}
\begin{equation}
\label{GRdef}
G_{R}(\eta) = \eta \tilde{K}_{R}(\eta),
\qquad
\tilde{K}_{R}(\eta) = K_{R}(4\eta\mmu^2),
\qquad
\eta = \frac{s}{4\mmu^2},
\qquad
s=q^2.
\rule{0mm}{7mm}
\end{equation}
Note that both real and imaginary parts of the ``spacelike'' hadronic vacuum
polarization function~$\bar\Pi(Q^{2})$ and the ``timelike'' kernel
function~$G_{R}\bigl(s/(4\mmu^2)\bigr)$~(\ref{GRdef}) contribute
to~Eq.~(\ref{THdef}).

In other words, in the case of the window function~(\ref{DefW2}) the
integration of~$F_{W_{2}}(q^2)$~(\ref{FWDef}) along the edges of its left cut
gives the left--hand side of Eq.~(\ref{RelPRint2}), the integration
of~$F_{W_{2}}(q^2)$ along the edges of its right cut yields the first term on
the right--hand side of Eq.~(\ref{RelPRint2}), whereas the integrals
of~$F_{W_{2}}(q^2)$ along the four semicircles of finite radii produce the
second term on the right--hand side of Eq.~(\ref{RelPRint2}), see
right--hand plot of Fig.~\ref{Plot:WF2}.

As~for the relation between the window quantities for~$a^{\text{HVP}}_{\mu}$
expressed in terms of functions~$\bar\Pi(Q^2)$ and~$D(Q^2)$, the integration
of Eq.~(\ref{AmuWD}) by parts yields
\begin{equation}
\label{KRelPDaux2}
A_{0}\!\int\limits_{Q_{1}^{2}}^{Q_{2}^{2}}\! D(Q^2) K_{D}(Q^2) \frac{d Q^2}{4\mmu^2} =
A_{0}\, \bar\Pi(Q^2) K_{D}(Q^2) \frac{Q^2}{4\mmu^2}\, \Biggr|_{Q_{1}^{2}}^{Q_{2}^{2}}
+ A_{0}\!\int\limits_{Q_{1}^{2}}^{Q_{2}^{2}}\! \bar\Pi(Q^2) K_{\Pi}(Q^2) \frac{d Q^2}{4\mmu^2},
\end{equation}
with Eqs.~(\ref{GDR_DP}) and~(\ref{KRelPD}) being employed. Contrary to the
previous case~(\ref{KRelPDaux1}), the first term on the right--hand side of
Eq.~(\ref{KRelPDaux2}) does not vanish, therefore (see Eq.~(43) of
Ref.~\cite{Nesterenko:2023bdc})
\begin{equation}
\label{RelPDint}
a^{\text{HVP}}_{\mu,\Pi,W_{2}}
=
a^{\text{HVP}}_{\mu,D,W_{2}}
+ \Delta a^{\text{HVP}}_{\mu,D,W_{2}},
\qquad
\Delta a^{\text{HVP}}_{\mu,D,W_{2}} =
- A_{0} \Bigl[ U(Q_{2}^{2}) - U(Q_{1}^{2}) \Bigr]\!,
\end{equation}
where
\begin{align}
\label{GDdef}
& U(Q^{2}) = \bar\Pi(Q^{2})G_{D}\biggl(\frac{Q^{2}}{4m_{\mu}^{2}}\biggr)\!,
\qquad
G_{D}(\zeta) = \zeta \tilde{K}_{D}(\zeta),
\qquad
\tilde{K}_{D}(\zeta) = K_{D}(4\zeta\mmu^2),
\nonumber \\[2mm]
& \zeta = \frac{Q^2}{4\mmu^2},
\qquad
Q^2=-q^2.
\end{align}

Thus, Eqs.~(\ref{RelPRint2}) and~(\ref{RelPDint}) imply that in the case of
the abrupt symmetric window function~$W_{2}(q^2)$~(\ref{DefW2}) the
representations~(\ref{AmuW}) are equivalent to each other only if the
additional contributions to~$a^{\text{HVP}}_{\mu,D,W_{2}}$~(\ref{AmuWD})
and~$a^{\text{HVP}}_{\mu,R,W_{2}}$~(\ref{AmuWR}) due to the window edge
effects are taken into account, specifically
\begin{equation}
\label{AmuW2}
a^{\text{HVP}}_{\mu,\Pi,W_{2}} =
a^{\text{HVP}}_{\mu,D,W_{2}} + \Delta a^{\text{HVP}}_{\mu,D,W_{2}} =
a^{\text{HVP}}_{\mu,R,W_{2}} + \Delta a^{\text{HVP}}_{\mu,R,W_{2}}.
\end{equation}

To~exemplify the obtained results, the window
quantities~$a^{\text{HVP}(\ell)}_{\mu,W_{n}}$~(\ref{AmuW}) will be evaluated
in the leading order of perturbation theory, i.e.,~in the second order in the
electromagnetic coupling~($\ell=2$). The~explicit expressions for the
required kernel functions (namely,~$\KG{\Pi}{2}(Q^2)$~\cite{Groote:2001vu,
Blum:2002ii, Nesterenko:2014txa, deRafael:2017gay},
$\KG{D}{2}(Q^2)$~\cite{Groote:2001vu, deRafael:2017gay},
and~$\KG{R}{2}(s)$~\cite{Berestetskii:1956tkr, Durand:1962zzb,
Brodsky:1967sr, Lautrup:1968tdb}) can all be found in~Sect.~2.2
of~Ref.~\cite{Nesterenko:2023bdc}. For~the window
function~$W_{2}(q^2)$~(\ref{DefW2}) the following values of parameters are
taken:
\begin{alignat}{2}
\label{WF2par}
Q_{1}^{2} &= 0.25\,\text{GeV}^2,
&\qquad
Q_{2}^{2} &= 0.50\,\text{GeV}^2,
\nonumber \\*[2mm]
s_{1} &= 0.25\,\text{GeV}^2,
&\qquad
s_{2} &= 0.50\,\text{GeV}^2.
\end{alignat}
As~mentioned earlier, the functions~$\bar\Pi(Q^2)$, $D(Q^2)$, and~$R(s)$
entering Eq.~(\ref{AmuW}) are calculated within~DPT~\cite{Nesterenko:2013vja,
Nesterenko:2014txa, Nesterenko:2016pmx}, and in Sect.~\ref{Sect:Results} the
number of active flavours is taken to be fixed~($\nf=3$) in the whole energy
range, that makes the DPT~expressions for the functions on hand continuous
(see~also Sect.~2.3 and Sect.~3.2 of Ref.~\cite{Nesterenko:2023bdc}).
Additionally, in what follows the fourth order in the strong coupling is
assumed (see~Chap.~4, Chap.~5, App.~B of Ref.~\cite{Nesterenko:2016pmx},
Sect.~5 of~Ref.~\cite{Nesterenko:2017wpb}, and Sect.~3.1 of
Ref.~\cite{Nesterenko:2019rag} for technical details) and the
PDG24~\cite{ParticleDataGroup:2024cfk} values of the involved Standard Model
parameters, including the world average~$\alpha_{s}(M_{Z}^{2})$,
are~employed.

Thus, the evaluation of the window
quantities~$a^{\text{HVP}(2)}_{\mu,W_{2}}$~(\ref{AmuW}) and the corresponding
additional contributions~$\Delta
a^{\text{HVP}(2)}_{\mu,D,W_{2}}$~(\ref{RelPDint}) and~$\Delta
a^{\text{HVP}(2)}_{\mu,R,W_{2}}$~(\ref{DltAmuRW2}) yields
\begin{align}
\label{AmuW2eval}
a^{\text{HVP}(2)}_{\mu,\Pi,W_{2}} & = (25.75 \pm 0.10) \!\times\! 10^{-10}, && \nonumber \\[2mm]
a^{\text{HVP}(2)}_{\mu,D,W_{2}}   & =  (7.85 \pm 0.02) \!\times\! 10^{-10}, &
\Delta a^{\text{HVP}(2)}_{\mu,D,W_{2}} & = (17.90 \pm 0.07) \!\times\! 10^{-10},  \\[2mm]
a^{\text{HVP}(2)}_{\mu,R,W_{2}}   & = (183.55 \pm 0.60) \!\times\! 10^{-10}, &
\Delta a^{\text{HVP}(2)}_{\mu,R,W_{2}} & = (-157.80 \pm 0.51) \!\times\! 10^{-10}. \nonumber
\end{align}
Factually, the equivalence relation~(\ref{AmuW2}) is confirmed by the mean
values given in Eq.~(\ref{AmuW2eval}), whereas the uncertainties (that, as
noted above, are caused by those of the involved Standard Model
parameters~\cite{ParticleDataGroup:2024cfk}) are specified to assess the
phenomenological impact.

\subsection{Abrupt asymmetric window function}
\label{Sect:WF3}

\begin{figure}[t]
\centerline{%
\includegraphics[height=70mm,clip]{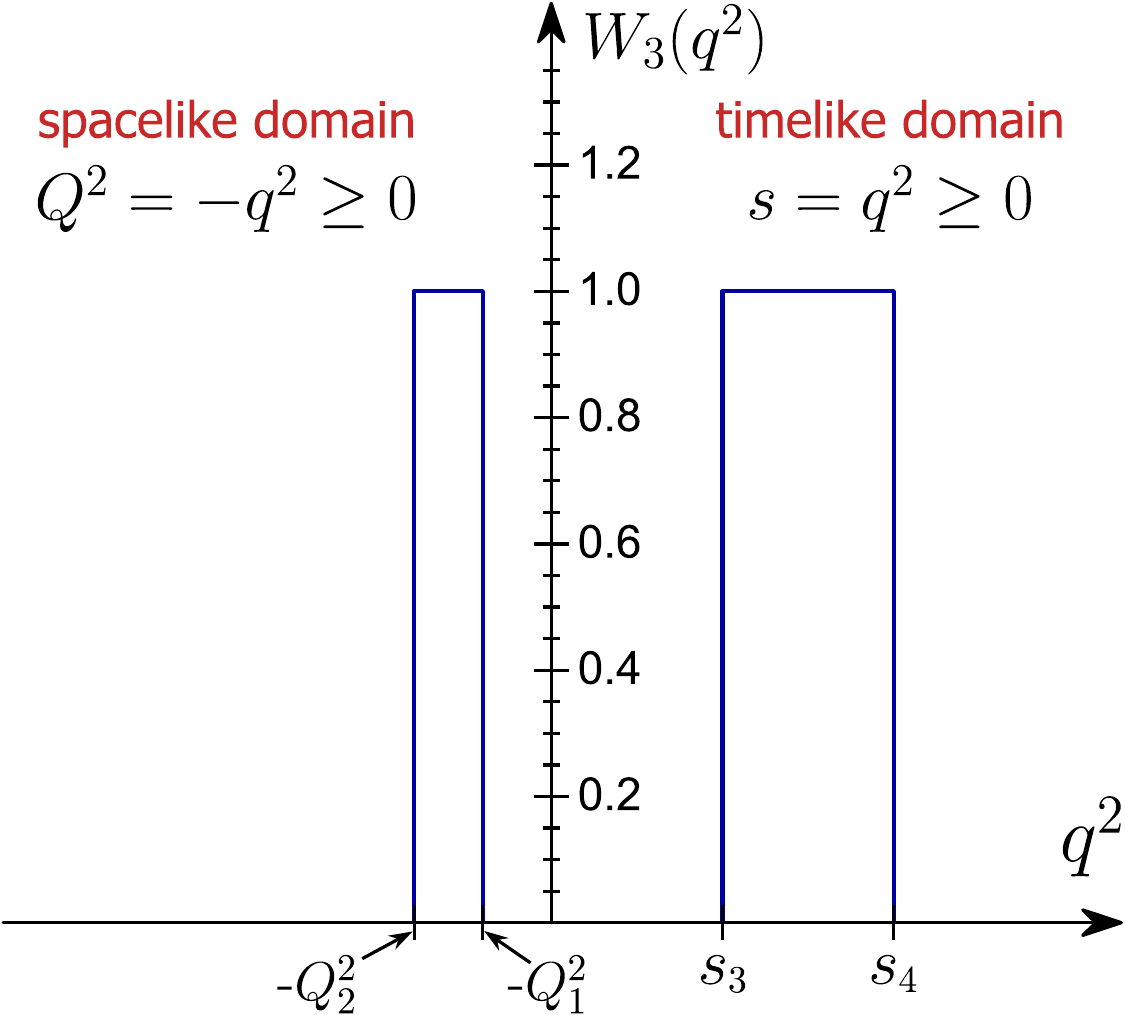}%
\hspace{10mm}%
\includegraphics[height=70mm,clip]{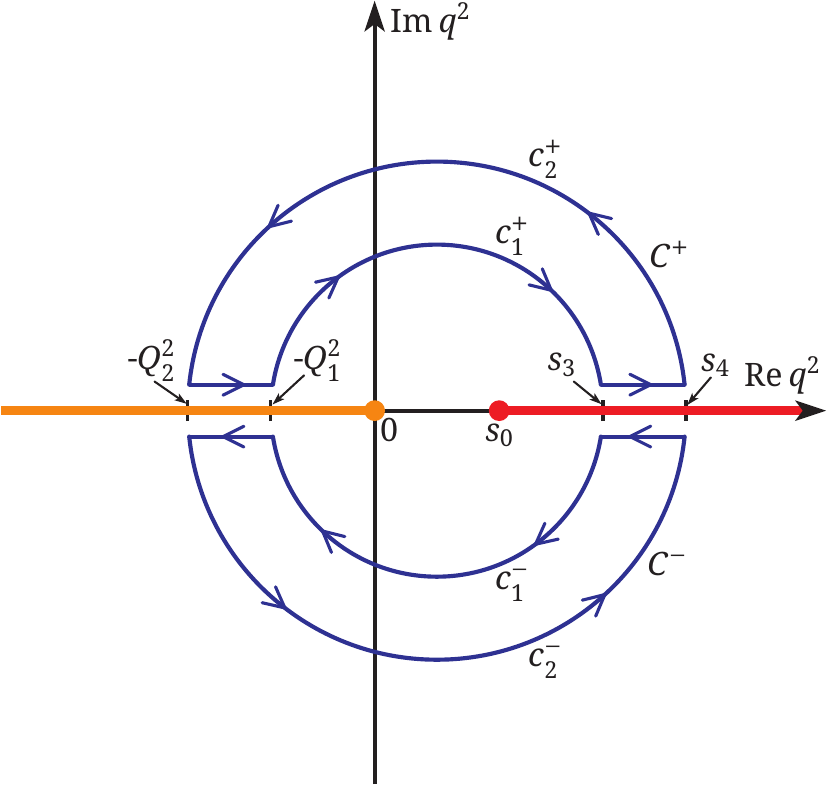}}
\caption{Left--hand plot: the abrupt asymmetric window
function~$W_{3}(q^2)$~(\ref{DefW3}). Right--hand plot: the closed integration
contours in the complex~$q^2$--plane in Eq.~(\ref{IntC1C2W3}).}
\label{Plot:WF3}
\end{figure}

Let us address now another kind of abrupt window function (see left--hand
plot of Fig.~\ref{Plot:WF3}):
\begin{align}
\label{DefW3}
W_{3}(q^2) & =
\theta(q^{2}+Q_{2}^{2}) - \theta(q^{2}+Q_{1}^{2}) +
\theta(q^{2}-s_{3}) - \theta(q^{2}-s_{4}),
\nonumber \\[2mm] & \quad\;
\left|-Q_{1}^{2}\right| \neq s_{3},
\qquad
\left|-Q_{2}^{2}\right| \neq s_{4},
\end{align}
where $\theta(x)$ is defined in Eq.~(\ref{DefUSF}) and it is assumed
hereinafter that the variables~$s_{i}$ ($i=3,4$) take positive values
and~\mbox{$s_{3} < s_{4}$}. The~function $W_{3}(q^{2})$~(\ref{DefW3}) covers
different kinematic intervals in spacelike ($Q_{1}^{2} \le Q^{2} \le
Q_{2}^{2}$) and timelike ($s_{3} \le s \le s_{4}$) domains and is hence
asymmetric with respect to the spacelike/timelike flip~$q^2 \leftrightarrow
-q^2$.

Similarly to the previous case, the integrals of the
function~$F_{W_{3}}(q^2)$~(\ref{FWDef}) along each of the closed
contours~$C^{+}$ and~$C^{-}$ displayed in the right--hand plot of
Fig.~\ref{Plot:WF3} vanish, therefore
\begin{equation}
\label{IntC1C2W3}
\oint_{C^{+}}\! F_{W_{3}}(q^2) d q^{2} +
\oint_{C^{-}}\! F_{W_{3}}(q^2) d q^{2} = 0,
\end{equation}
that ultimately yields
\begin{equation}
\label{RelPRint1W3}
A_{0}\!\int\limits_{Q_{1}^{2}}^{Q_{2}^{2}} \!
\bar\Pi(Q^2) K_{\Pi}(Q^2) \frac{dQ^2}{4\mmu^2} =
A_{0}\!\int\limits_{s_{3}}^{s_{4}} \!
R(s) K_{R}(s) \frac{ds}{4\mmu^2}
- A_{0}\frac{\Delta F_{3}}{2\pi i},
\end{equation}
where
\begin{equation}
\label{DltFW3}
\Delta F_{3} =
\int_{c_{1}^{+}}\! F_{W_{3}}(q^2) d q^{2} +
\int_{c_{1}^{-}}\! F_{W_{3}}(q^2) d q^{2} +
\int_{c_{2}^{+}}\! F_{W_{3}}(q^2) d q^{2} +
\int_{c_{2}^{-}}\! F_{W_{3}}(q^2) d q^{2}.
\end{equation}
Then, it is convenient to make the following change of the integration
variables in Eq.~(\ref{DltFW3}) (\mbox{$j=1$}~for the first two terms and
\mbox{$j=2$}~for the last two terms):
\begin{align}
\label{PolCoordsW3}
q^2 = q^{2}_{C,j} + q^{2}_{R,j} e^{i\varphi},
\qquad &
q^{2}_{C,1} = \frac{1}{2}\Bigl(s_{3}-\left|-Q^{2}_{1}\right|\Bigr),
\qquad
q^{2}_{R,1} = \frac{1}{2}\Bigl(s_{3}+\left|-Q^{2}_{1}\right|\Bigr),
\nonumber \\[1mm]
\qquad &
q^{2}_{C,2} = \frac{1}{2}\Bigl(s_{4}-\left|-Q^{2}_{2}\right|\Bigr),
\qquad
q^{2}_{R,2} = \frac{1}{2}\Bigl(s_{4}+\left|-Q^{2}_{2}\right|\Bigr).
\end{align}
For example, the first term on the right--hand side of Eq.~(\ref{DltFW3})
takes the~form
\begin{equation}
\label{DltFW3c1}
\int_{c_{1}^{+}}\! F_{W_{3}}(q^2) d q^{2} =
i \!\int\limits_{\pi-\varepsilon}^{\varepsilon}
\!\bar H(q_{C,1}^{2},q_{R,1}^{2},\varphi) d\varphi,
\end{equation}
where [cf.~Eq.~(\ref{THdef})]
\begin{equation}
\label{HBdef}
\bar H(q_{C}^{2},q_{R}^{2},\varphi) =
\bar\Pi\Bigl[-\bigl(q^{2}_{C} + q^{2}_{R} e^{i\varphi}\bigr)\Bigr]
G_{R}\!\left(\frac{q^{2}_{C} + q^{2}_{R} e^{i\varphi}}{4\mmu^2}\right)\!\!
\Biggl(1 + \frac{q^{2}_{C}}{q^{2}_{R}}e^{-i\varphi}\Biggr)^{\! -1}
\end{equation}
and~$G_{R}(\eta)$ is given in Eq.~(\ref{GRdef}).
Thus,~Eq.~(\ref{RelPRint1W3}) eventually leads to the relation between the
spacelike~(\ref{AmuWP}) and timelike~(\ref{AmuWR}) window quantities:
\begin{equation}
\label{RelPRint3}
a^{\text{HVP}}_{\mu,\Pi,W_{3}} =
a^{\text{HVP}}_{\mu,R,W_{3}} +
\Delta a^{\text{HVP}}_{\mu,R,W_{3}},
\end{equation}
where
\begin{equation}
\label{DltAmuRW3}
\Delta a^{\text{HVP}}_{\mu,R,W_{3}} =
- A_{0}\Bigl[ \bar T(q_{C,2}^{2},q_{R,2}^{2}) - \bar T(q_{C,1}^{2},q_{R,1}^{2}) \Bigr]\!,
\end{equation}
\begin{equation}
\label{TBdef}
\bar T(q_{C}^{2},q_{R}^{2}) = \frac{1}{2\pi}\!
\left[ \int\limits_{\varepsilon}^{\pi-\varepsilon}\!
\bar H(q_{C}^{2},q_{R}^{2},\varphi) d\varphi
+
\int\limits_{\pi+\varepsilon}^{2\pi-\varepsilon}\!
\bar H(q_{C}^{2},q_{R}^{2},\varphi) d\varphi \right]\!\!,
\end{equation}
and~$\bar H(q_{C}^{2},q_{R}^{2},\varphi)$ is defined in Eq.~(\ref{HBdef}).

As~in the preceding Section, for the window function~(\ref{DefW3}) the
integration of~$F_{W_{3}}(q^2)$~(\ref{FWDef}) along the edges of its left cut
yields the left--hand side of Eq.~(\ref{RelPRint3}), the integration
of~$F_{W_{3}}(q^2)$ along the edges of its right cut gives the first term on
the right--hand side of Eq.~(\ref{RelPRint3}), while the second term on the
right--hand side of Eq.~(\ref{RelPRint3}) is generated by the integrals
of~$F_{W_{3}}(q^2)$ along the four semicircles of finite radii, see
right--hand plot of Fig.~\ref{Plot:WF3}.

As~for the relation between the window quantities~(\ref{AmuWP})
and~(\ref{AmuWD}), it evidently retains the same form as in the previous
case:
\begin{equation}
\label{RelPDintW3}
a^{\text{HVP}}_{\mu,\Pi,W_{3}}
=
a^{\text{HVP}}_{\mu,D,W_{3}}
+ \Delta a^{\text{HVP}}_{\mu,D,W_{3}},
\qquad
\Delta a^{\text{HVP}}_{\mu,D,W_{3}} =
\Delta a^{\text{HVP}}_{\mu,D,W_{2}},
\end{equation}
where $\Delta a^{\text{HVP}}_{\mu,D,W_{2}}$ is given in Eq.~(\ref{RelPDint}).

Therefore, Eqs.~(\ref{RelPRint3}) and~(\ref{RelPDintW3}) indicate that for
the abrupt asymmetric window function~$W_{3}(q^2)$~(\ref{DefW3}) the
representations for~$a^{\text{HVP}}_{\mu,W_{3}}$~(\ref{AmuW}) are mutually
equivalent only if the additional contributions
to~$a^{\text{HVP}}_{\mu,D,W_{3}}$~(\ref{AmuWD})
and~$a^{\text{HVP}}_{\mu,R,W_{3}}$~(\ref{AmuWR}) produced by the window edge
effects are included, namely
\begin{equation}
\label{AmuW3}
a^{\text{HVP}}_{\mu,\Pi,W_{3}} =
a^{\text{HVP}}_{\mu,D,W_{3}} + \Delta a^{\text{HVP}}_{\mu,D,W_{3}} =
a^{\text{HVP}}_{\mu,R,W_{3}} + \Delta a^{\text{HVP}}_{\mu,R,W_{3}}.
\end{equation}

To~illustrate the obtained results, the following values of parameters
are taken for the window function~$W_{3}(q^2)$~(\ref{DefW3}):
\begin{alignat}{2}
\label{WF3par}
Q_{1}^{2} &= 0.10\,\text{GeV}^2,
&\qquad
Q_{2}^{2} &= 0.20\,\text{GeV}^2,
\nonumber \\*[2mm]
s_{3} &= 0.25\,\text{GeV}^2,
&\qquad
s_{4} &= 0.50\,\text{GeV}^2,
\end{alignat}
that, in turn, leads~to
\begin{align}
\label{AmuW3eval}
a^{\text{HVP}(2)}_{\mu,\Pi,W_{3}} & = (69.63 \pm 0.28) \!\times\! 10^{-10}, &&\nonumber \\[2mm]
a^{\text{HVP}(2)}_{\mu,D,W_{3}}   & = (28.39 \pm 0.10) \!\times\! 10^{-10},&
\Delta a^{\text{HVP}(2)}_{\mu,D,W_{3}} & = (41.24 \pm 0.18) \!\times\! 10^{-10},  \\[2mm]
a^{\text{HVP}(2)}_{\mu,R,W_{3}}   & = (183.55 \pm 0.61) \!\times\! 10^{-10},&
\Delta a^{\text{HVP}(2)}_{\mu,R,W_{3}} & = (-113.92 \pm 0.32) \!\times\! 10^{-10}, \nonumber
\end{align}
and affirms the equivalence relation~(\ref{AmuW3}).

\subsection{Smooth symmetric window function}
\label{Sect:WF4}

\begin{figure}[t]
\centerline{%
\includegraphics[height=70mm,clip]{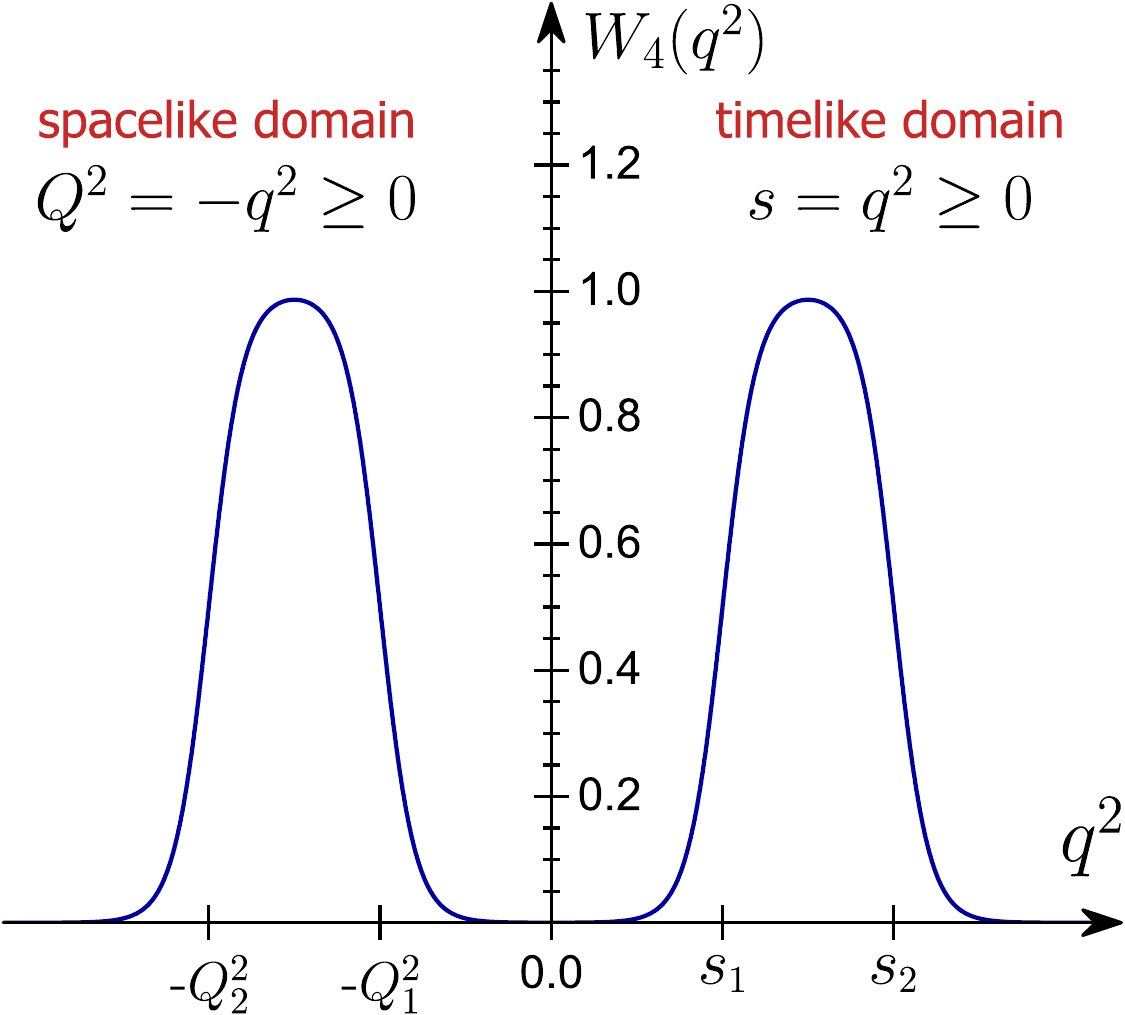}%
\hspace{10mm}%
\includegraphics[height=70mm,clip]{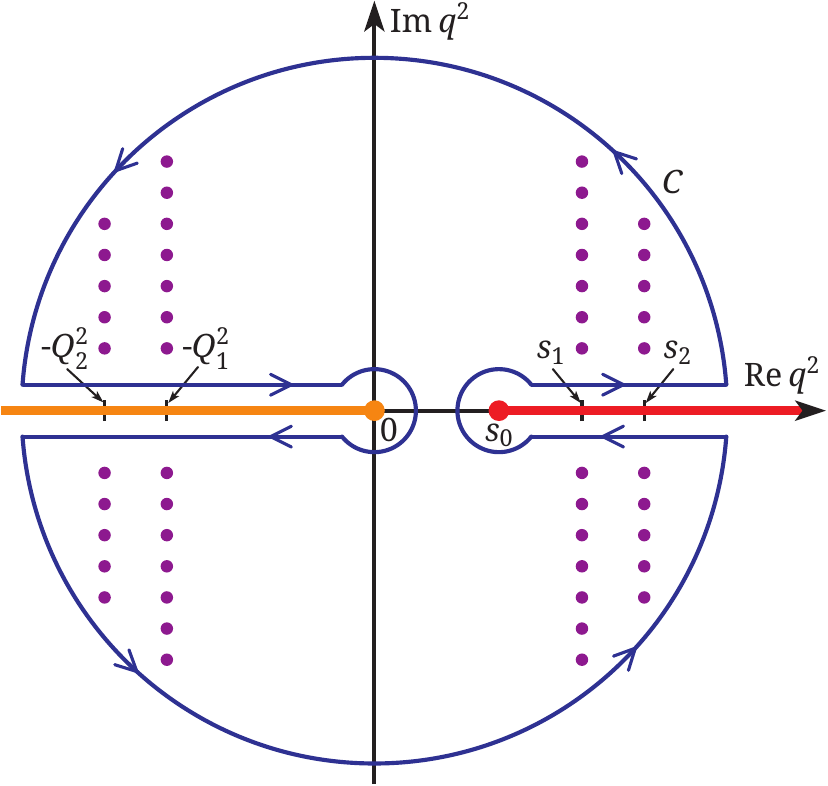}}
\caption{Left--hand plot: the smooth symmetric window
function~$W_{4}(q^2)$~(\ref{DefW4}). Right--hand plot: the closed integration
contour in the complex~$q^2$--plane in Eq.~(\ref{IntW4}).}
\label{Plot:WF4}
\end{figure}

Let us turn now to the smooth window function similar to that of proposed in
Ref.~\cite{RBC:2018dos} (see left--hand plot of Fig.~\ref{Plot:WF4}):
\begin{align}
\label{DefW4}
W_{4}(q^{2}) & =
\Theta(q^{2}+Q_{2}^{2}, \Delta_{\text{SL}}) -
\Theta(q^{2}+Q_{1}^{2}, \Delta_{\text{SL}}) +
\Theta(q^{2}-s_{1}, \Delta_{\text{TL}}) -
\Theta(q^{2}-s_{2}, \Delta_{\text{TL}}),
\nonumber \\*[2mm] & \quad\;
\left|-Q_{1}^{2}\right| = s_{1},
\qquad
\left|-Q_{2}^{2}\right| = s_{2},
\qquad
\Delta_{\text{SL}} = \Delta_{\text{TL}},
\end{align}
where
\begin{equation}
\label{DefUSFS}
\Theta(q^{2},\Delta) =
\frac{1}{2}\biggl[1+\tanh\biggl(\frac{q^{2}}{\Delta}\biggr)\!\biggr] \!.
\end{equation}
The hyperbolic tangent~$\tanh(x)$ entering this equation has an infinite
number of the first order poles in the complex $x$--plane (which will be
shown by the purple dots in the contour plots hereinafter) located at the
points~$x = \pm\,i\pi(n+1/2)$, with~$n$ being a non--negative integer
(\mbox{$n = 0, 1, 2, ...$}). The~window function $W_{4}(q^{2})$~(\ref{DefW4})
secludes the coequal kinematic ranges in spacelike and timelike domains and
is therefore symmetric with respect to the spacelike/timelike flip~\mbox{$q^2
\leftrightarrow -q^2$}.

Contrary to the abrupt window function (Sect.~\ref{Sect:WF2} and
Sect.~\ref{Sect:WF3}), the smooth window
function~$W_{4}(q^{2})$~(\ref{DefW4}) does not affect the integration limits
in Eq.~(\ref{AmuW}), but instead brings an infinite number of isolated poles
into the complex~$q^2$--plane.
Namely, the integral of the function~$F_{W_{4}}(q^2)$~(\ref{FWDef}) along the
closed contour~$C$ shown in the right--hand plot of Fig.~\ref{Plot:WF4}
does not vanish
\begin{equation}
\label{IntW4}
\frac{1}{2\pi i}\oint_{C} F_{W_{4}}(q^2) d q^{2} =
\sum_{\text{poles}} \text{Res}\, F_{W_{4}}(q^2).
\end{equation}
The~integration on the left--hand side of this equation, which can be
performed in the same way as that of Sect.~\ref{Sect:WF1}, eventually yields
the difference between the window quantities~(\ref{AmuWP}) and~(\ref{AmuWR})
\begin{equation}
\label{IntW4-1}
\frac{A_{0}}{2\pi i}\oint_{C} F_{W_{4}}(q^2) d q^{2} =
a^{\text{HVP}}_{\mu,\Pi,W_{4}} -
a^{\text{HVP}}_{\mu,R,W_{4}}.
\end{equation}
In~turn, the right--hand side of Eq.~(\ref{IntW4}) can be represented~as
\begin{align}
\label{IntW4res}
& \Delta a^{\text{HVP}}_{\mu,R,W_{4}} =
A_{0}\!\sum_{\text{poles}} \text{Res}\, F_{W_{4}}(q^2) =
\nonumber \\*[1mm] = \, &
A_{0\!}\sum_{n=0}^{\infty}
\biggl[
  V(-Q_{2}^{2},\Delta_{\text{SL}},n)
- V(-Q_{1}^{2},\Delta_{\text{SL}},n)
+ V(s_{1},\Delta_{\text{TL}},n)
- V(s_{2},\Delta_{\text{TL}},n)
\biggr]\!,
\end{align}
where
\begin{equation}
\label{Vdef}
V(q^{2},\Delta,n) =
\frac{\Delta}{2}
\Biggl\{\!
F_{W_{1}}\!\biggl[q^{2}+i\pi\Delta\biggl(\frac{1}{2}+n\biggr)\biggr]\!
+
F_{W_{1}}\!\biggl[q^{2}-i\pi\Delta\biggl(\frac{1}{2}+n\biggr)\biggr]\!
\Biggr\}
\end{equation}
and the function $F_{W_{1}}(q^2)$ is defined in Eqs.~(\ref{FWDef})
and~(\ref{DefW1}). Thus,~Eqs.~(\ref{IntW4})--(\ref{IntW4res}) result~in the
following relation between the spacelike~(\ref{AmuWP}) and
timelike~(\ref{AmuWR}) window quantities:
\begin{equation}
\label{RelPRint4}
a^{\text{HVP}}_{\mu,\Pi,W_{4}} =
a^{\text{HVP}}_{\mu,R,W_{4}} +
\Delta a^{\text{HVP}}_{\mu,R,W_{4}}.
\end{equation}

Similarly to the previous cases, for the window function~(\ref{DefW4}) the
integration of~$F_{W_{4}}(q^2)$~(\ref{FWDef}) along the edges of its left cut
gives the left--hand side of Eq.~(\ref{RelPRint4}), the integration
of~$F_{W_{4}}(q^2)$ along the edges of its right cut yields the first term on
the right--hand side of Eq.~(\ref{RelPRint4}), the~integrals
of~$F_{W_{4}}(q^2)$ along the circles of infinitely large and infinitely
small radii vanish, while the second term on the right--hand side of
Eq.~(\ref{RelPRint4}) is produced by the first order poles of the window
function~$W_{4}(q^{2})$~(\ref{DefW4}), see right--hand plot of
Fig.~\ref{Plot:WF4}.

As~earlier, the relation between the window quantities
for~$a^{\text{HVP}}_{\mu}$ expressed in terms of the functions~$\bar\Pi(Q^2)$
and~$D(Q^2)$ can be derived by integration of Eq.~(\ref{AmuWD}) by parts,
that leads~to
\begin{align}
\label{KRelPDW4}
& A_{0}\!\int\limits_{0}^{\infty}\! D(Q^2) K_{D}(Q^2) W_{4}(-Q^2) \frac{d Q^2}{4\mmu^2}
= A_{0}\, \bar\Pi(Q^2) K_{D}(Q^2) W_{4}(-Q^2) \frac{Q^2}{4\mmu^2}\, \Biggr|_{0}^{\infty} +
\nonumber \\[1.25mm]
+ \, & A_{0}\!\int\limits_{0}^{\infty}\! \bar\Pi(Q^2) K_{\Pi}(Q^2) W_{4}(-Q^2) \frac{d Q^2}{4\mmu^2}
- A_{0}\!\int\limits_{0}^{\infty}\! \bar\Pi(Q^2) K_{D}(Q^2) \widetilde{W}_{4}(Q^2) \frac{d Q^2}{4\mmu^2},
\end{align}
with Eqs.~(\ref{GDR_DP}) and~(\ref{KRelPD}) being employed. In~this equation
\begin{align}
\label{WT4def}
\widetilde{W}_{4}(Q^2) & = \frac{d\,W_{4}(-Q^2)}{d\,\ln Q^2} =
\nonumber \\*[2mm] & \hspace{-10mm}
= Q^{2} \Bigl[
-\widetilde{\Theta}\bigl(Q^{2}-Q_{2}^{2},\Delta_{\text{SL}}\bigr)
+\widetilde{\Theta}\bigl(Q^{2}-Q_{1}^{2},\Delta_{\text{SL}}\bigr)
-\widetilde{\Theta}\bigl(Q^{2}+s_{1},    \Delta_{\text{TL}}\bigr)
+\widetilde{\Theta}\bigl(Q^{2}+s_{2},    \Delta_{\text{TL}}\bigr)
\Bigr]\!,
\rule[-5mm]{0mm}{5mm}
\end{align}
where
\begin{equation}
\widetilde{\Theta}\bigl(Q^{2},\Delta\bigr) =
\biggl[2 \Delta \cosh^{2}\biggl(\frac{Q^{2}}{\Delta}\biggr)\biggr]^{-1}.
\end{equation}
Then, since the first term on the right--hand side of Eq.~(\ref{KRelPDW4})
vanishes (see~Ref.~\cite{Nesterenko:2021byp}), one arrives~at
\begin{equation}
\label{RelPDW4}
a^{\text{HVP}}_{\mu,\Pi,W_{4}} =
a^{\text{HVP}}_{\mu,D,W_{4}} + \Delta a^{\text{HVP}}_{\mu,D,W_{4}},
\qquad
\Delta a^{\text{HVP}}_{\mu,D,W_{4}} =
A_{0}\!\int\limits_{0}^{\infty}\! \bar\Pi(Q^2) K_{D}(Q^2) \widetilde{W}_{4}(Q^2) \frac{d Q^2}{4\mmu^2},
\end{equation}
where~$\widetilde{W}_{4}(Q^2)$ is defined in Eq.~(\ref{WT4def}).

Therefore, Eqs.~(\ref{RelPRint4}) and~(\ref{RelPDW4}) indicate that for the
smooth symmetric window function~$W_{4}(q^2)$~(\ref{DefW4}) the
representations for~$a^{\text{HVP}}_{\mu,W_{4}}$~(\ref{AmuW}) are mutually
equivalent only if the additional contributions
to~$a^{\text{HVP}}_{\mu,D,W_{4}}$~(\ref{AmuWD})
and~$a^{\text{HVP}}_{\mu,R,W_{4}}$~(\ref{AmuWR}), which appear due to the
window edge effects, are taken into account, specifically
\begin{equation}
\label{AmuW4}
a^{\text{HVP}}_{\mu,\Pi,W_{4}} =
a^{\text{HVP}}_{\mu,D,W_{4}} + \Delta a^{\text{HVP}}_{\mu,D,W_{4}} =
a^{\text{HVP}}_{\mu,R,W_{4}} + \Delta a^{\text{HVP}}_{\mu,R,W_{4}}.
\end{equation}

To~exemplify the obtained results, the following values of parameters
are taken for the window function~$W_{4}(q^2)$~(\ref{DefW4}):
\begin{alignat}{3}
\label{WF4par}
Q_{1}^{2} &= 0.25\,\text{GeV}^2,
&\qquad
Q_{2}^{2} &= 0.50\,\text{GeV}^2,
&\qquad
\Delta_{\text{SL}} &= 0.05\,\text{GeV}^2,
\nonumber \\[2mm]
s_{1} &= 0.25\,\text{GeV}^2,
&\qquad
s_{2} &= 0.50\,\text{GeV}^2,
&\qquad
\Delta_{\text{TL}} &= 0.05\,\text{GeV}^2,
\end{alignat}
that eventually yields
\begin{align}
\label{AmuW4eval}
a^{\text{HVP}(2)}_{\mu,\Pi,W_{4}} & = (27.86 \pm 0.10) \!\times\! 10^{-10}, && \nonumber \\[2mm]
a^{\text{HVP}(2)}_{\mu,D,W_{4}}   & = (8.77 \pm 0.03) \!\times\! 10^{-10}, &
\Delta a^{\text{HVP}(2)}_{\mu,D,W_{4}} & = (19.09 \pm 0.08) \!\times\! 10^{-10}, \\[2mm]
a^{\text{HVP}(2)}_{\mu,R,W_{4}}   & = (189.19 \pm 0.64) \!\times\! 10^{-10}, &
\Delta a^{\text{HVP}(2)}_{\mu,R,W_{4}} & = (-161.33 \pm 0.54) \!\times\! 10^{-10}, \nonumber
\end{align}
thereby confirming the equivalence relation~(\ref{AmuW4}).

\subsection{Smooth asymmetric window function}
\label{Sect:WF5}

\begin{figure}[t]
\centerline{%
\includegraphics[height=70mm,clip]{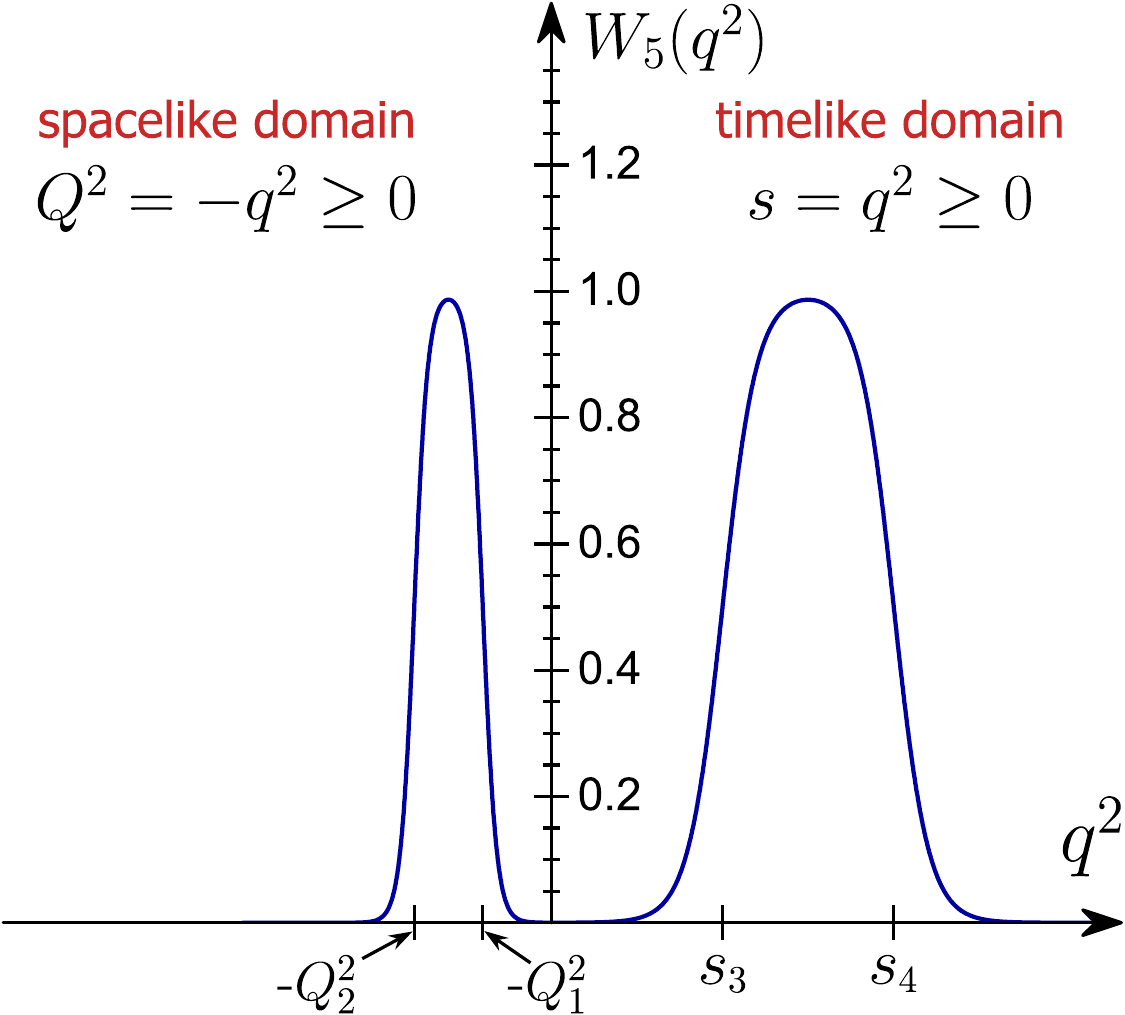}%
\hspace{10mm}%
\includegraphics[height=70mm,clip]{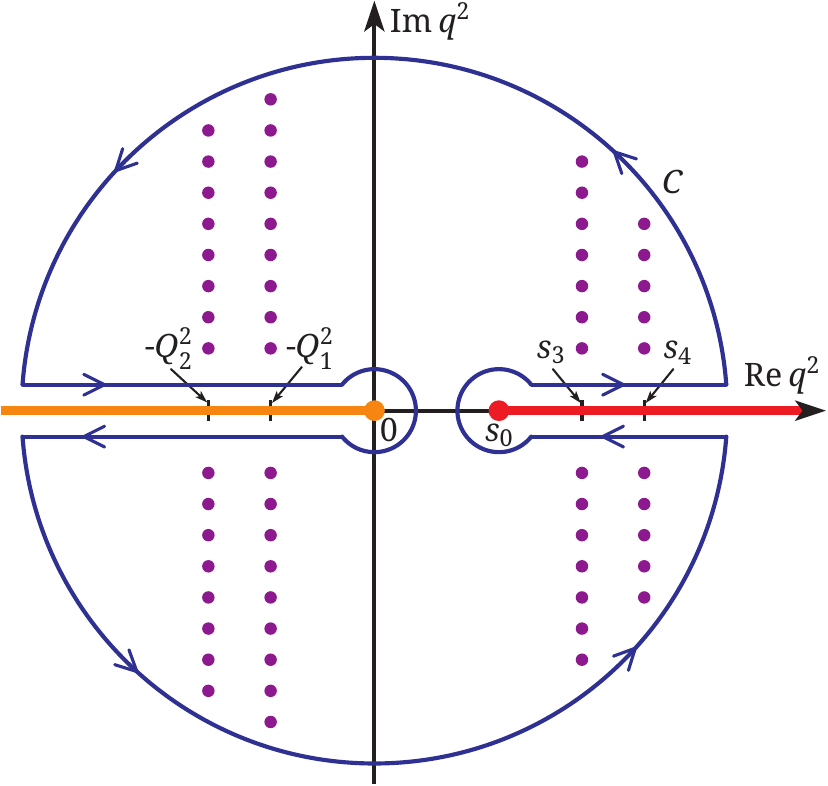}}
\caption{Left--hand plot: the smooth asymmetric window
function~$W_{5}(q^2)$~(\ref{DefW5}). Right--hand plot: the closed integration
contour in the complex~$q^2$--plane in Eq.~(\ref{IntW5}).}
\label{Plot:WF5}
\end{figure}

Finally, let us address another type of smooth window function (see
left--hand plot of Fig.~\ref{Plot:WF5}):
\begin{align}
\label{DefW5}
W_{5}(q^2) & =
\Theta(q^{2}+Q_{2}^{2}, \Delta_{\text{SL}}) -
\Theta(q^{2}+Q_{1}^{2}, \Delta_{\text{SL}}) +
\Theta(q^{2}-s_{3}, \Delta_{\text{TL}}) -
\Theta(q^{2}-s_{4}, \Delta_{\text{TL}}),
\nonumber \\[2mm] & \quad\;
\left|-Q_{1}^{2}\right| \neq s_{3},
\qquad
\left|-Q_{2}^{2}\right| \neq s_{4},
\end{align}
where $\Theta(q^{2},\Delta)$ is defined in Eq.~(\ref{DefUSFS}). The~function
$W_{5}(q^{2})$~(\ref{DefW5}) secludes different kinematic ranges in spacelike
and timelike domains and is hence asymmetric with respect to the
spacelike/timelike flip~$q^2 \leftrightarrow -q^2$.

Factually, all the calculations made for the smooth symmetric window
function~(\ref{DefW4}) in~Sect.~\ref{Sect:WF4} remain valid also for the
asymmetric case~(\ref{DefW5}). Specifically, starting~from
\begin{equation}
\label{IntW5}
\frac{1}{2\pi i}\oint_{C} F_{W_{5}}(q^2) d q^{2} =
\sum_{\text{poles}} \text{Res}\, F_{W_{5}}(q^2)
\end{equation}
(the~corresponding closed integration contour~$C$ is displayed in the
right--hand plot of Fig.~\ref{Plot:WF5}) and following the same lines as
above, one ultimately arrives~at
\begin{align}
\label{AmuW5}
& a^{\text{HVP}}_{\mu,\Pi,W_{5}} =
a^{\text{HVP}}_{\mu,D,W_{5}} + \Delta a^{\text{HVP}}_{\mu,D,W_{5}} =
a^{\text{HVP}}_{\mu,R,W_{5}} + \Delta a^{\text{HVP}}_{\mu,R,W_{5}},
\nonumber \\[2mm] &
\Delta a^{\text{HVP}}_{\mu,D,W_{5}} = \Delta a^{\text{HVP}}_{\mu,D,W_{4}},
\qquad
\Delta a^{\text{HVP}}_{\mu,R,W_{5}} = \Delta a^{\text{HVP}}_{\mu,R,W_{4}},
\end{align}
i.e.,~the representations for~$a^{\text{HVP}}_{\mu,W_{5}}$~(\ref{AmuW}) are
equivalent to each other only if the additional contributions
to~$a^{\text{HVP}}_{\mu,D,W_{5}}$~(\ref{AmuWD})
and~$a^{\text{HVP}}_{\mu,R,W_{5}}$~(\ref{AmuWR}) generated by the window edge
effects [see~Eqs.~(\ref{RelPDW4}) and~(\ref{IntW4res}), respectively]
are~included. Here it is assumed that in Eqs.~(\ref{RelPDW4})
and~(\ref{IntW4res}) variables $s_{1}$ and~$s_{2}$ are replaced with $s_{3}$
and~$s_{4}$, respectively.

To~illustrate these results, the following values of parameters are taken for
the window function~$W_{5}(q^2)$~(\ref{DefW5}):
\begin{alignat}{3}
\label{WF5par}
Q_{1}^{2} &= 0.10\,\text{GeV}^2,
&\qquad
Q_{2}^{2} &= 0.20\,\text{GeV}^2,
&\qquad
\Delta_{\text{SL}} &= 0.02\,\text{GeV}^2,
\nonumber \\*[2mm]
s_{3} &= 0.25\,\text{GeV}^2,
&\qquad
s_{4} &= 0.50\,\text{GeV}^2,
&\qquad
\Delta_{\text{TL}} &= 0.05\,\text{GeV}^2,
\end{alignat}
that leads~to
\begin{align}
\label{AmuW5eval}
a^{\text{HVP}(2)}_{\mu,\Pi,W_{5}} & = (73.32 \pm 0.30) \!\times\! 10^{-10}, && \nonumber \\[2mm]
a^{\text{HVP}(2)}_{\mu,D,W_{5}}   & = (30.60 \pm 0.11) \!\times\! 10^{-10}, &
\Delta a^{\text{HVP}(2)}_{\mu,D,W_{5}} & = (42.72 \pm 0.19) \!\times\! 10^{-10}, \\[2mm]
a^{\text{HVP}(2)}_{\mu,R,W_{5}}   & = (189.19 \pm 0.64) \!\times\! 10^{-10}, &
\Delta a^{\text{HVP}(2)}_{\mu,R,W_{5}} & = (-115.87 \pm 0.34) \!\times\! 10^{-10}, \nonumber
\end{align}
thereby reaffirming the equivalence relation~(\ref{AmuW5}).

\section{Discussion}
\label{Sect:Dsc}

\subsection{General comments}
\label{Sect:GC}

First of all, it is worthwhile to note that the results obtained in
Sect.~\ref{Sect:WF3} constitute the generalization of those of
Sect.~\ref{Sect:WF2}. In~particular, the expression for~$\Delta
a^{\text{HVP}}_{\mu,R,W_{3}}$~(\ref{DltAmuRW3}) is~clearly applicable for
arbitrary values of~$Q_{i}^{2}$ and~$s_{i}$, that shape the abrupt asymmetric
window function~$W_{3}(q^2)$~(\ref{DefW3}). For~example,
Eq.~(\ref{DltAmuRW3}) remains valid also for the cases
\mbox{$\bigl\{\left|-Q_{1}^{2}\right| = s_{3},$} \mbox{$\left|-Q_{2}^{2}\right| \neq s_{4}\bigr\}$,}
\mbox{$\bigl\{\left|-Q_{1}^{2}\right| \neq s_{3}, \left|-Q_{2}^{2}\right| = s_{4}\bigr\}$},
as~well~as
\mbox{$\bigl\{\left|-Q_{1}^{2}\right| = s_{3}, \left|-Q_{2}^{2}\right| = s_{4}\bigr\}$},
which corresponds to the symmetric window function~$W_{2}(q^2)$~(\ref{DefW2}).
Additionally, the results obtained in Sect.~\ref{Sect:WF4}
and~Sect.~\ref{Sect:WF5} are also applicable for arbitrary values
of~$Q_{i}^{2}$, $s_{i}$, $\Delta_{\text{SL}}$, and~$\Delta_{\text{TL}}$
in~each of four terms on the right--hand sides of Eqs.~(\ref{DefW4})
and~(\ref{DefW5}), that shape the respective smooth window functions.

Summarizing the obtained results, it is necessary to emphasize that the
window quantities for HVP~contributions to the muon anomalous magnetic
moment~$a^{\text{HVP}}_{\mu,W}$~(\ref{AmuW}) expressed in~terms of the
hadronic vacuum polarization function, the Adler function, and the
\mbox{$R$--ratio} of electron--positron annihilation into hadrons are
equivalent to each other only if the additional terms appearing due to the
window edge effects are properly taken into account, namely
\begin{equation}
\label{AmuWn}
a^{\text{HVP}}_{\mu,\Pi,W_{n}} =
a^{\text{HVP}}_{\mu,D,W_{n}} + \Delta a^{\text{HVP}}_{\mu,D,W_{n}} =
a^{\text{HVP}}_{\mu,R,W_{n}} + \Delta a^{\text{HVP}}_{\mu,R,W_{n}}.
\end{equation}
While in the absence of integrand's modulation $\Delta
a^{\text{HVP}}_{\mu,D,W_{1}}=0$ and~$\Delta a^{\text{HVP}}_{\mu,R,W_{1}}=0$
(Sect.~\ref{Sect:WF1}), in the case of abrupt
(Sects.~\ref{Sect:WF2},~\ref{Sect:WF3}) and smooth
(Sects.~\ref{Sect:WF4},~\ref{Sect:WF5}) window functions such additional
contributions become nonvanishing and have to be taken into account.
The~corresponding explicit expressions for $\Delta
a^{\text{HVP}}_{\mu,D,W_{n}}$ and~$\Delta a^{\text{HVP}}_{\mu,R,W_{n}}$ are
gathered in Tab.~\ref{Tab:WEFs}.

\subsection{Hybrid evaluation of~$a^{\text{\normalfont{HVP}}}_{\mu}$}
\label{Sect:AmuHbd}

The~results presented in Sects.~\ref{Sect:WF2} and~\ref{Sect:WF3} factually
enable one to evaluate~$a^{\text{HVP}}_{\mu}$ by~making simultaneous use of
the inputs for functions~$\bar\Pi(Q^2)$, $D(Q^2)$, and~$R(s)$ at~various
energies. Specifically, in the first line of Eq.~(\ref{W1IntC3}) the
corresponding kinematic range can be decomposed into~$N+1$~intervals
$(Q_{0}^{2}=0, Q_{N+1}^{2}=\infty)$
\begin{equation}
\label{Amu2Hbd0}
a^{\text{HVP}}_{\mu} = A_{0}\!\int\limits_{0}^{\infty}\! \bar\Pi(Q^2)
K_{\Pi}(Q^2) \frac{d Q^2}{4\mmu^2} =
A_{0}\sum_{i=0}^{N}\int\limits_{Q_{i}^2}^{Q_{i+1}^2}\! \bar\Pi(Q^2)
K_{\Pi}(Q^2) \frac{d Q^2}{4\mmu^2}
\end{equation}
and then an individual integral on the right--hand side of this equation can
be expressed in~terms of~$D(Q^2)$ or~$R(s)$ by applying Eq.~(\ref{RelPDint})
or Eq.~(\ref{RelPRint3}), respectively.

To~exemplify such a hybrid assessment of~$a^{\text{HVP}}_{\mu}$, let~us
address the canonical form of the leading--order (i.e.,~the second order
in~the electromagnetic coupling) hadronic vacuum polarization contribution to
the muon anomalous magnetic moment (see~Sect.~2.2
of~Ref.~\cite{Nesterenko:2023bdc} and~references therein)
\begin{equation}
\label{Amu2Hbd1}
a^{\text{HVP(2)}}_{\mu} = a^{\text{HVP(2)}}_{\mu,\Pi} =
a^{\text{HVP(2)}}_{\mu,\Pi,W_{1}} =
\APF{2}\!\!\int\limits_{0}^{\infty}\!
\bar\Pi(Q^2) \KG{\Pi}{2}(Q^2) \frac{d Q^2}{4\mmu^2} =
(695.95 \pm 2.83) \!\times\! 10^{-10}.
\end{equation}
The~numerical evaluation of~the quantities in this Section is~performed
by~making use of the four--loop DPT~\cite{Nesterenko:2013vja,
Nesterenko:2014txa, Nesterenko:2016pmx} and employs
the~PDG24~\cite{ParticleDataGroup:2024cfk} values of the involved Standard
Model parameters, see also Sect.~\ref{Sect:AmuDPT}.

In~Eq.~(\ref{Amu2Hbd1}) the complete integration range can be split into
three intervals, that corresponds to $N=2$ in Eq.~(\ref{Amu2Hbd0}):
\begin{equation}
\label{Amu2Hbd2}
a^{\text{HVP(2)}}_{\mu} =
a^{\text{HVP(2)}}_{\mu,\Pi,W_{2}}(0,Q_{1}^{2}) +
a^{\text{HVP(2)}}_{\mu,\Pi,W_{2}}(Q_{1}^{2},Q_{2}^{2}) +
a^{\text{HVP(2)}}_{\mu,\Pi,W_{2}}(Q_{2}^{2},\infty).
\end{equation}
The second and third terms on the right--hand side of Eq.~(\ref{Amu2Hbd2})
can then be expressed in~terms of the \mbox{$R$--ratio} of electron--positron
annihilation into hadrons and the Adler function by~making use of,
respectively, Eqs.~(\ref{RelPRint2}) and~(\ref{RelPDint}):
\begin{align}
\label{Amu2Hbd3}
a^{\text{HVP(2)}}_{\mu} & =
a^{\text{HVP(2)}}_{\mu,\Pi,W_{2}}(0,Q_{1}^{2}) +
\nonumber \\[2mm] & +
a^{\text{HVP(2)}}_{\mu,R,W_{2}}(s_{1},s_{2}) +
\Delta a^{\text{HVP(2)}}_{\mu,R,W_{2}}(s_{1},s_{2}) +
\\[2mm] & +
a^{\text{HVP(2)}}_{\mu,D,W_{2}}(Q_{2}^{2},\infty) +
\Delta a^{\text{HVP(2)}}_{\mu,D,W_{2}}(Q_{2}^{2},\infty),
\qquad
\left|-Q_{1}^{2}\right| = s_{1},
\quad
\left|-Q_{2}^{2}\right| = s_{2}.
\nonumber
\end{align}
The functions appearing on the right--hand sides of Eqs.~(\ref{Amu2Hbd2})
and~(\ref{Amu2Hbd3}) constitute the spacelike and timelike window quantities
studied in Sect.~\ref{Sect:WF2}, while their arguments (being omitted
in~Sect.~\ref{Sect:WF2} for the sake of simplicity) specify the parameters of
the abrupt symmetric window function~(\ref{DefW2}):
\begin{align}
\label{Amu2Hbd4}
& a^{\text{HVP}}_{\mu,\Pi,W_{2}}(Q_{1}^{2},Q_{2}^{2}) =
A_{0}\!\int\limits_{Q_{1}^{2}}^{Q_{2}^{2}} \!
\bar\Pi(Q^2) K_{\Pi}(Q^2) \frac{dQ^2}{4\mmu^2}, &&
\\[1.5mm]
\label{Amu2Hbd5}
& a^{\text{HVP}}_{\mu,D,W_{2}}(Q_{1}^{2},Q_{2}^{2}) =
A_{0}\!\int\limits_{Q_{1}^{2}}^{Q_{2}^{2}}\! D(Q^2) K_{D}(Q^2) \frac{d Q^2}{4\mmu^2}, &
\quad\!\!
& \Delta a^{\text{HVP}}_{\mu,D,W_{2}}(Q_{1}^{2},Q_{2}^{2}) =
- A_{0} \Bigl[ U(Q_{2}^{2}) - U(Q_{1}^{2}) \Bigr]\!,
\\[-1.5mm]
\label{Amu2Hbd6}
& a^{\text{HVP}}_{\mu,R,W_{2}}(s_{1},s_{2})  =
A_{0}\!\int\limits_{s_{1}}^{s_{2}} \! R(s) K_{R}(s) \frac{ds}{4\mmu^2}, &
\quad
& \Delta a^{\text{HVP}}_{\mu,R,W_{2}}(s_{1},s_{2}) =
- A_{0}\Bigl[ T(s_{2}) - T(s_{1}) \Bigr]\!,
\end{align}
where the functions~$U(Q^{2})$ and~$T(s)$ are given in, respectively,
Eqs.~(\ref{GDdef}) and~(\ref{THdef}), see Sect.~\ref{Sect:WF2} for the
details. It~is worthwhile to note here that in the second line of
Eq.~(\ref{Amu2Hbd3}) the timelike integration range can also be taken
different from~$\left|-Q_{1}^{2}\right| \le s \le \left|-Q_{2}^{2}\right|$.
In~the latter case the contribution due to the spacelike/timelike asymmetric
window edge effects~(\ref{DltAmuRW3}) should be used instead of the symmetric
one~(\ref{Amu2Hbd6}), see Sect.~\ref{Sect:WF3} for the details.

As~an example, the following values of the window parameters entering
Eq.~(\ref{Amu2Hbd3}) are employed
\begin{alignat}{2}
\label{WFHbd}
Q_{1}^{2} &= 0.25\,\text{GeV}^2,
&\qquad
Q_{2}^{2} &= 0.50\,\text{GeV}^2,
\nonumber \\*[2mm]
s_{1} &= 0.25\,\text{GeV}^2,
&\qquad
s_{2} &= 0.50\,\text{GeV}^2,
\end{alignat}
that leads~to
\begin{align}
\label{Amu2HbdEval}
a^{\text{HVP}(2)}_{\mu,\Pi,W_{2}}(0,Q_{1}^{2}) & = (654.54 \pm 2.69) \!\times\! 10^{-10}, && \nonumber \\[2mm]
a^{\text{HVP}(2)}_{\mu,R,W_{2}}(s_{1},s_{2})   & = (183.55 \pm 0.60) \!\times\! 10^{-10}, &
\Delta a^{\text{HVP}(2)}_{\mu,R,W_{2}}(s_{1},s_{2}) & = (-157.80 \pm 0.51) \!\times\! 10^{-10}, \nonumber \\[2mm]
a^{\text{HVP}(2)}_{\mu,D,W_{2}}(Q_{2}^{2},\infty)   & = (353.53 \pm 0.86) \!\times\! 10^{-12}, &
\Delta a^{\text{HVP}(2)}_{\mu,D,W_{2}}(Q_{2}^{2},\infty) & = (119.88 \pm 0.43) \!\times\! 10^{-11}.
\end{align}
As~discussed in Sect.~3 of Ref.~\cite{Nesterenko:2023bdc}, in~the presence of
the quark flavour thresholds the DPT functions~$\bar\Pi(Q^2)$, $D(Q^2)$,
and~$R(s)$ (likewise their perturbative counterparts) are piecewise
continuous, that eventually generates additional contributions to
Eq.~(\ref{Amu2Hbd3}), namely
\begin{equation}
\label{DltPDHQT}
\Delta a_{\mu,[4]}^{D^{\!},(2)} = (134.21 \pm 1.72) \!\times\! 10^{-13},
\qquad\;\;\,
\Delta a_{\mu,[5]}^{D^{\!},(2)} = (455.28 \pm 3.11) \!\times\! 10^{-16},
\end{equation}
see Ref.~\cite{Nesterenko:2023bdc} for the details. Thus, the mean values
specified in Eqs.~(\ref{Amu2HbdEval}) and~(\ref{DltPDHQT}) confirm the
equivalence between the canonical~(\ref{Amu2Hbd1}) and
hybrid~(\ref{Amu2Hbd3}) expressions for the hadronic vacuum polarization
contribution to the muon anomalous magnetic moment, while the uncertainties
are given to assess the phenomenological impact.

\subsection{Updated DPT value of~$a_{\mu}$}
\label{Sect:AmuDPT}

This section reports an update of the assessment~\cite{Nesterenko:2014txa,
Nesterenko:2015rea, Nesterenko:2017hqv, Nesterenko:2023bdc} of the hadronic
vacuum polarization contributions to the muon anomalous magnetic moment
performed within dispersively improved perturbation
theory~\cite{Nesterenko:2013vja, Nesterenko:2014txa, Nesterenko:2016pmx}.
In~particular, as~discussed in Refs.~\cite{Nesterenko:2014txa,
Nesterenko:2015rea, Nesterenko:2017hqv, Nesterenko:2023bdc}, the~DPT~has
proved to be capable of describing~$a^{\text{HVP}}_{\mu}$ without using
timelike data on~\mbox{$R$--ratio}. Specifically, the applicability of
DPT~hadronic vacuum polarization function in the whole energy range makes it
possible to directly employ it in the corresponding expression
for~$a^{\text{HVP}}_{\mu}$ (see,~respectively, Eqs.~(22a) and~(8a) of
Ref.~\cite{Nesterenko:2023bdc}). Additionally, the DPT hadronic vacuum
polarization function can provide a complementing infrared input to the MUonE
experiment~\cite{CarloniCalame:2015obs, MUonE:2016hru, MUonE:2019qlm,
Banerjee:2020tdt} in the energy range uncovered by measurements, see
Ref.~\cite{Nesterenko:2023bdc} for the details.

\begin{figure}[t]
\centerline{\includegraphics[width=110mm,clip]{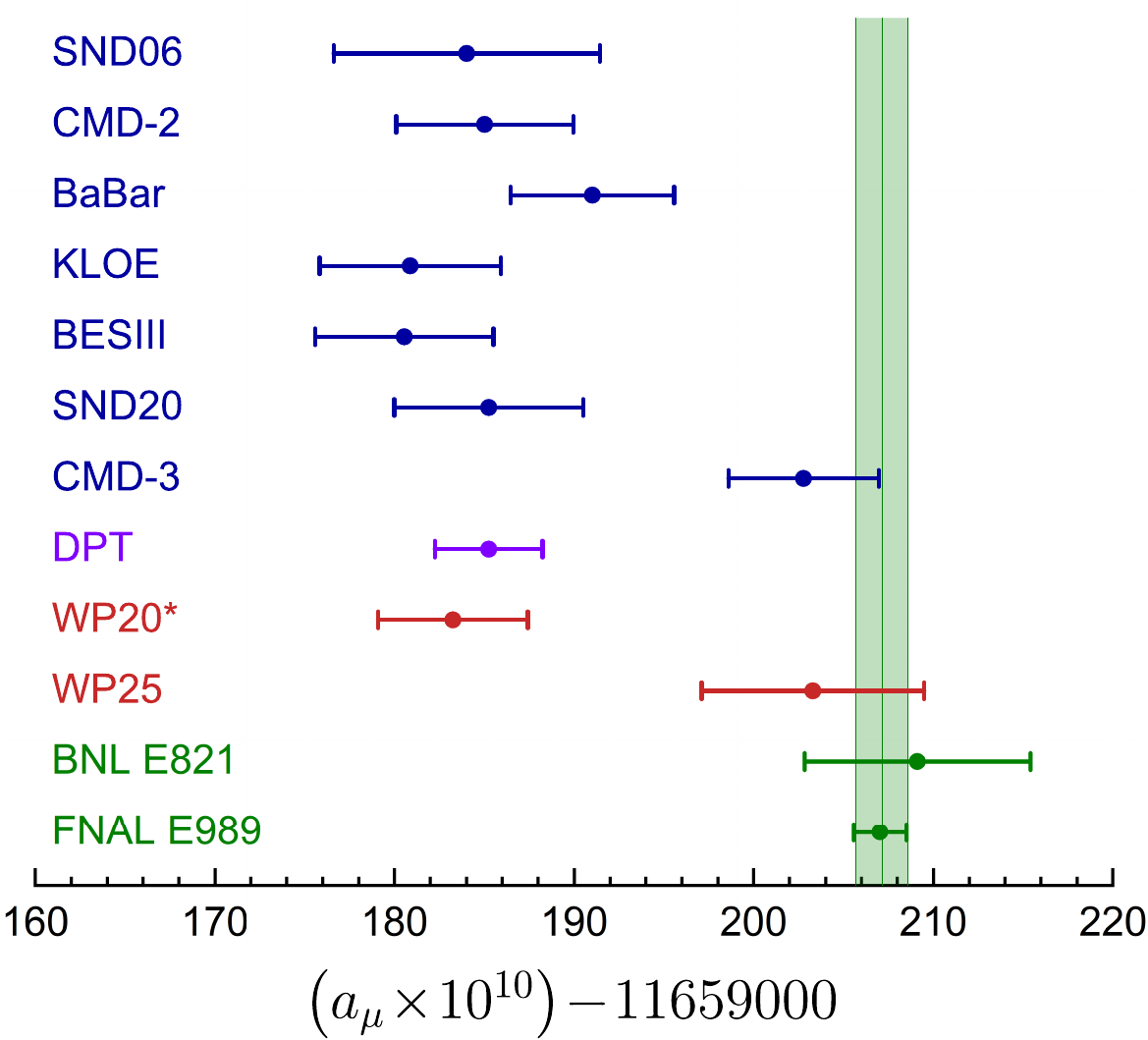}}
\caption{A~summary of the muon anomalous magnetic moment
assessments~\cite{Aliberti:2025beg} (labeled SND06, CMD--2, BaBar, KLOE,
\mbox{BESIII}, SND20, CMD--3, WP20$^{*}$, and~WP25), its
evaluation~(\ref{AmuSMDPT25})~(labeled~DPT), experimental
measurements~\cite{Muong-2:2006rrc} and~\cite{Muong-2:2021ojo,
Muong-2:2023cdq, Muong-2:2025xyk} (labeled, respectively, BNL~E821 and
FNAL~E989), and their~combined result~(\ref{AmuSMexp25}) (vertical shaded
band), see~main text for the details.}
\label{Plot:Amu25}
\end{figure}

Thus, in~the leading and in the next--to--leading orders of perturbation
theory [i.e.,~in~the second~($\ell=2$) and in~the third~($\ell=3$) orders
in~the electromagnetic coupling] the DPT hadronic vacuum polarization
function leads~to [as~earlier, the~fourth order in the strong coupling
is~assumed and the PDG24~\cite{ParticleDataGroup:2024cfk}~values of the
involved Standard Model parameters, including the world
average~$\alpha_{s}(M_{Z}^{2})$, are~used]
\begin{equation}
\label{Amu2DPT25}
\amu{2} = (695.95 \pm 2.83)\!\times\! 10^{-10},
\end{equation}
\begin{equation}
\label{Amu3DPT25}
\begin{split}
\amu{3a} &= (-218.08 \pm 0.96)\!\times\! 10^{-11},
\quad\,
a^{\text{HVP}(3b)}_{\mu, e} = (106.53 \pm 0.44)\!\times\! 10^{-11},
\\[2mm]
a^{\text{HVP}(3b)}_{\mu, \tau} &= (559.67 \pm 2.08)\!\times\! 10^{-15},
\qquad
\amu{3c} = (334.92 \pm 2.66)\!\times\! 10^{-13},
\end{split}
\end{equation}
that agrees with data--driven evaluations~\cite{Aliberti:2025beg,
Davier:2019can, Keshavarzi:2019abf, Kurz:2014wya, Jegerlehner:2017lbd}.
In~Eqs.~(\ref{Amu2DPT25}) and~(\ref{Amu3DPT25}) the term~``$(2)$''
corresponds to the diagram with single hadronic insertion, the
term~``$(3a)$'' refers to the diagrams with either an~additional photon line
with all possible permutations or an~additional muon loop insertion, the
terms~``$(3b)$'' embody the contributions of the diagrams with an additional
electron loop or $\tau$--lepton loop insertions, whereas the term~``$(3c)$''
corresponds to the diagram with double hadronic insertion, see
Refs.~\cite{Nesterenko:2021byp, Nesterenko:2023bdc} and references therein
for the details. The values~(\ref{Amu2DPT25}) and~(\ref{Amu3DPT25}), being
supplemented with the remaining contributions~$a^{\text{QED}}_{\mu}$,
$a^{\text{EW}}_{\mu}$, $\amu{4}$, $a^{\text{HLbL}}_{\mu}$ reported~in
Ref.~\cite{Aliberti:2025beg}, yield
\begin{equation}
\label{AmuSMDPT25}
a^{\text{SM}}_{\mu} = (11659185.25 \pm 3.00)\!\times\! 10^{-10},
\end{equation}
that constitutes an update of the evaluations performed in
Refs.~\cite{Nesterenko:2014txa, Nesterenko:2015rea, Nesterenko:2017hqv,
Nesterenko:2023bdc}.

A summary of the muon anomalous magnetic moment assessments reported in
WP25~\cite{Aliberti:2025beg} is displayed in Fig.~\ref{Plot:Amu25}. Each of
the seven blue datapoints (labeled SND06, CMD--2, BaBar, KLOE, \mbox{BESIII},
SND20, CMD--3) comprises all the contributions to~$a_{\mu}$ taken from
WP25~\cite{Aliberti:2025beg}, except for~$\amu{2}$, which is evaluated in the
following way. The corresponding measurements of \mbox{$e^{+}e^{-} \to
\pi^{+}\pi^{-}$} channel below~$1.8$~GeV were treated within CHKLS, DHMZ, and
KNTW approaches (see,~respectively, Sect.~2.6.1, Sect.~2.5.1, and Sect.~2.5.2
of Ref.~\cite{Aliberti:2025beg}) and then the averaged outcome was
complemented with the contributions from other channels and from the energies
exceeding~$1.8$~GeV taken from WP20~\cite{Aoyama:2020ynm}, see Sect.~2
and~Sect.~9 of WP25~\cite{Aliberti:2025beg}. The~measurements SND06, CMD--2,
BaBar, KLOE, \mbox{BESIII} entered the result for WP20~\cite{Aoyama:2020ynm}
(again, the red datapoint labeled WP20$^{*}$ combines~$\amu{2}$ from
WP20~\cite{Aoyama:2020ynm} with the other contributions to~$a_{\mu}$ from
WP25~\cite{Aliberti:2025beg}), while SND20 and CMD--3 appeared afterwards.
Due~to yet unresolved tension between CMD--3 and the other experimental
measurements, the data--driven contribution~$\amu{2}$ was not included in the
WP25 assessment~of~$a_{\mu}$ (the red datapoint labeled~WP25), which uses the
lattice evaluation of~$\amu{2}$ instead, see WP25~\cite{Aliberti:2025beg} and
references therein for the details. The~measurements of~$a_{\mu}$ by
BNL~E821~\cite{Muong-2:2006rrc} and FNAL~E989~\cite{Muong-2:2021ojo,
Muong-2:2023cdq, Muong-2:2025xyk} experiments are shown by the green
datapoints, while their~combined result
\begin{equation}
\label{AmuSMexp25}
a^{\text{exp}}_{\mu} = (11659207.15 \pm 1.45)\!\times\! 10^{-10}
\end{equation}
is represented by the vertical shaded band. As~one can infer from
Fig.~\ref{Plot:Amu25}, the DPT~assessment~(\ref{AmuSMDPT25}) (the violet
datapoint labeled~DPT) agrees with most of the data--driven evaluations
of~$a_{\mu}$ and differs by $6.6$~standard deviations from its experimental
value~(\ref{AmuSMexp25}).

\section{Conclusions}
\label{Sect:Concl}

The relations between the window quantities for the hadronic vacuum
polarization contributions to the muon anomalous magnetic moment in spacelike
and timelike domains are~studied. Two~types of window functions [namely,
abrupt~(Sects.~\ref{Sect:WF2} and~\ref{Sect:WF3}) and smooth
(Sects.~\ref{Sect:WF4} and~\ref{Sect:WF5})], as well as two kinds of
kinematic intervals [namely, symmetric~(Sects.~\ref{Sect:WF2}
and~\ref{Sect:WF4}) and asymmetric~(Sects.~\ref{Sect:WF3} and~\ref{Sect:WF5})
with respect to the spacelike/timelike flip] are~addressed. It~is shown that
the window quantities for~$a^{\text{HVP}}_{\mu}$ represented in~terms of the
hadronic vacuum polarization function, the Adler function, and the $R$--ratio
of electron--positron annihilation into hadrons are mutually equivalent only
if the additional contributions due to the window edge effects are properly
taken into account and the explicit expressions for such contributions are
derived, see Eqs.~(\ref{AmuW2}), (\ref{AmuW3}), (\ref{AmuW4}),
(\ref{AmuW5}),~(\ref{AmuWn}) and~Tab.~\ref{Tab:WEFs}. The~obtained results
enable one to evaluate~$a^{\text{HVP}}_{\mu}$ by~making simultaneous use of
the inputs for functions~$\bar\Pi(Q^2)$, $D(Q^2)$, and~$R(s)$ at~various
energies and an example of such hybrid assessment is provided. The~obtained
results also enable one to accurately compare the window quantities
for~$a^{\text{HVP}}_{\mu}$ based, e.g., on MUonE or lattice data with the
ones based on $R$--ratio data, even~if the window function covers different
kinematic ranges in spacelike and timelike domains.

\begin{landscape}

\begin{table}[t]
\small%
\vspace*{-7.5mm}
\caption{Additional contributions to~$a^{\text{HVP}}_{\mu,W}$~(\ref{AmuWn})
due to the window edge effects.}
\label{Tab:WEFs}
\vskip2.5mm
\centering
\begin{tabular}{p{123.25mm}p{123.25mm}}
\hline\hline
Abrupt window function (Sects.~\ref{Sect:WF2},~~\ref{Sect:WF3})
\rule[-3.5mm]{0mm}{9.5mm}%
& Smooth window function (Sects.~\ref{Sect:WF4},~\ref{Sect:WF5}) \\
\hline
$\ds
W_{3}(q^2) =
\theta(q^{2}+Q_{2}^{2}) - \theta(q^{2}+Q_{1}^{2}) +
\theta(q^{2}-s_{3})     - \theta(q^{2}-s_{4})
\rule[-4mm]{0mm}{10mm}%
$
&
$\ds
W_{5}(q^2) =
\Theta(q^{2}+Q_{2}^{2}, \Delta_{2}) -
\Theta(q^{2}+Q_{1}^{2}, \Delta_{1}) +
\Theta(q^{2}-s_{3}, \Delta_{3}) -
\Theta(q^{2}-s_{4}, \Delta_{4})
$
\\ \hline
$\ds
\Delta a^{\text{HVP}}_{\mu,D,W_{3}} =
- A_{0}\Biggl[
\bar\Pi(Q_{2}^{2})G_{D}\biggl(\frac{Q_{2}^{2}}{4m_{\mu}^{2}}\biggr)\! -
\bar\Pi(Q_{1}^{2})G_{D}\biggl(\frac{Q_{1}^{2}}{4m_{\mu}^{2}}\biggr)\!
\Biggr],
$

\vskip3mm

$\ds
G_{D}(\zeta) = \zeta \tilde{K}_{D}(\zeta),
\qquad
\tilde{K}_{D}(\zeta) = K_{D}(4\zeta\mmu^2),
\qquad
\zeta = \frac{Q^2}{4\mmu^2}
$

&
$\ds
\Delta a^{\text{HVP}}_{\mu,D,W_{5}} =
A_{0}\!\int\limits_{0}^{\infty}\! \bar\Pi(Q^2) K_{D}(Q^2)
\widetilde{W}_{5}(Q^2) \frac{d Q^2}{4\mmu^2},
$
\rule{0mm}{10mm}%

\vskip2mm

$\ds%
\begin{aligned}
\widetilde{W}_{5}(Q^2) = \frac{d\,W_{5}(-Q^2)}{d\,\ln Q^2} & = Q^{2} \Bigl[
-\widetilde{\Theta}\bigl(Q^{2}-Q_{2}^{2},\Delta_{2}\bigr)
+\widetilde{\Theta}\bigl(Q^{2}-Q_{1}^{2},\Delta_{1}\bigr)
\\[0.5mm] &
-\widetilde{\Theta}\bigl(Q^{2}+s_{3},    \Delta_{3}\bigr)
+\widetilde{\Theta}\bigl(Q^{2}+s_{4},    \Delta_{4}\bigr)
\Bigr]
\rule[-4.5mm]{0mm}{4.5mm}%
\end{aligned}
$
\\ \hline
$\ds
\Delta a^{\text{HVP}}_{\mu,R,W_{3}} =
- A_{0}\Bigl[ \bar T(q_{C,2}^{2},q_{R,2}^{2}) - \bar T(q_{C,1}^{2},q_{R,1}^{2}) \Bigr],
$

\vskip3mm

$\ds
\bar T(q_{C}^{2},q_{R}^{2}) = \frac{1}{2\pi}\!
\left[ \int\limits_{\varepsilon}^{\pi-\varepsilon}\!
\bar H(q_{C}^{2},q_{R}^{2},\varphi) d\varphi
+
\int\limits_{\pi+\varepsilon}^{2\pi-\varepsilon}\!
\bar H(q_{C}^{2},q_{R}^{2},\varphi) d\varphi \right]\!\!,
$

\vskip2mm

$\ds
\bar H(q_{C}^{2},q_{R}^{2},\varphi) =
\bar\Pi\Bigl[-\bigl(q^{2}_{C} + q^{2}_{R} e^{i\varphi}\bigr)\Bigr]
G_{R}\!\left(\frac{q^{2}_{C} + q^{2}_{R} e^{i\varphi}}{4\mmu^2}\right)\!\!
\Biggl(1 + \frac{q^{2}_{C}}{q^{2}_{R}}e^{-i\varphi}\Biggr)^{\! -1}\!,
$

\vskip2.25mm

$\ds
q^{2}_{C,1} = \frac{1}{2}\Bigl(s_{3}-\left|-Q^{2}_{1}\right|\Bigr),
\quad
q^{2}_{R,1} = \frac{1}{2}\Bigl(s_{3}+\left|-Q^{2}_{1}\right|\Bigr),
$

\vskip2.75mm

$\ds
q^{2}_{C,2} = \frac{1}{2}\Bigl(s_{4}-\left|-Q^{2}_{2}\right|\Bigr),
\quad
q^{2}_{R,2} = \frac{1}{2}\Bigl(s_{4}+\left|-Q^{2}_{2}\right|\Bigr),
$

\vskip2.25mm

$\ds
G_{R}(\eta) = \eta \tilde{K}_{R}(\eta),
\qquad
\tilde{K}_{R}(\eta) = K_{R}(4\eta\mmu^2),
\qquad
\eta = \frac{s}{4\mmu^2}
$

\vspace*{4.5mm}

&

$\ds%
\Delta a^{\text{HVP}}_{\mu,R,W_{5}} =
A_{0\!}\sum_{n=0}^{\infty}
\biggl[
  V(-Q_{2}^{2},\Delta_{2},n)
- V(-Q_{1}^{2},\Delta_{1},n)
$
\rule{0mm}{9mm}%

$\ds%
\phantom{\Delta a^{\text{HVP}}_{\mu,R,W_{5}}}
+ V(s_{3},\Delta_{3},n)
- V(s_{4},\Delta_{4},n)
\biggr],
$

\vskip1.5mm

$\ds%
V(q^{2},\Delta,n) =
\frac{\Delta}{2}
\Biggl\{\!
F_{W_{1}}\!\biggl[q^{2}+i\pi\Delta\biggl(\frac{1}{2}+n\biggr)\biggr]\!
+
F_{W_{1}}\!\biggl[q^{2}-i\pi\Delta\biggl(\frac{1}{2}+n\biggr)\biggr]\!
\Biggr\},
$

\vskip2.5mm

$\ds%
F_{W_{1}}(q^2) =
\frac{1}{4\mmu^2} \bar\Pi(-q^2) K_{R}(q^2)
$

\\ \hline

\vskip-1mm

\parbox[t]{110mm}{%
Note that the window function $W_{3}(q^2)$ becomes symmetric with respect to
the spacelike/timelike flip~$q^2 \leftrightarrow -q^2$, when two conditions
$\left|-Q_{1}^{2}\right| = s_{3}$,
$\left|-Q_{2}^{2}\right| = s_{4}$
are fulfilled}

\vspace*{7mm}

&

\vskip-1.5mm

\parbox[t]{110mm}{%
Note that the window function $W_{5}(q^2)$ becomes symmetric with respect to
the spacelike/timelike flip~$q^2 \leftrightarrow -q^2$, when four conditions
$\left|-Q_{1}^{2}\right| = s_{3}$, $\Delta_{1} = \Delta_{3}$,
$\left|-Q_{2}^{2}\right| = s_{4}$, $\Delta_{2} = \Delta_{4}$
are fulfilled}

\\
\hline\hline
\end{tabular}
\end{table}

\end{landscape}

\end{document}